\documentclass{nws}

\usepackage{lineno,hyperref}
\usepackage{dirtytalk}

\usepackage{amsfonts, amssymb}
\usepackage{graphicx}
\usepackage{graphics}
\usepackage{multirow}
\usepackage{makecell}
\expandafter\let\csname equation*\endcsname\relax

\expandafter\let\csname endequation*\endcsname\relax
\usepackage{amsmath} 
\DeclareMathOperator*{\argmin}{arg\,min}
\DeclareMathOperator*{\argmax}{arg\,max}
\newcommand\round[1]{\big[#1\big]}

\title[Structural Analysis of Real and Model-Generated Networks]
      {Network Classification Based Structural Analysis of Real Networks and their Model-Generated Counterparts}

 \author[M. Nagy and R. Molontay]
        {MARCELL NAGY\\
         Department of Stochastics, Budapest University of Technology and Economics, Budapest, Hungary \\
         \and~ROLAND MOLONTAY \\
         MTA-BME Stochastics Research Group, Budapest, Hungary
         \email{marcessz@math.bme.hu, molontay@math.bme.hu}}

\begin{document}

\label{firstpage}

\maketitle

\begin{abstract}
Data-driven analysis of complex networks has been in the focus of research for decades. An important area of research is to study how well real networks can be described with a small selection of metrics, furthermore how well network models can capture the relations between graph metrics observed in real networks. In this paper, we apply machine learning techniques to investigate the aforementioned problems. We study 500 real-world networks along with 2,000 synthetic networks generated by four frequently used network models with previously calibrated parameters to make the generated graphs as similar to the real networks as possible.
This paper unifies several branches of data-driven complex network analysis, such as the study of graph metrics and their pair-wise relationships, network similarity estimation, model calibration, and graph classification.
We find that the correlation profiles of the structural measures significantly differ across network domains and the domain can be efficiently determined using a small selection of graph metrics. The structural properties of the network models with fixed parameters are robust enough to perform parameter calibration. The goodness-of-fit of the network models highly depends on the network domain. 
By solving classification problems, we find that the models lack the capability of generating a graph with a high clustering coefficient and relatively large diameter simultaneously. On the other hand, models are able to capture exactly the degree-distribution-related metrics.
\end{abstract}


\section{Introduction}
Data-based analysis of complex networks has attracted a lot of research interest since the millennium when the prompt evolution of information technology made the comprehensive exploration of real networks possible. The study of networks pervades the majority of disciplines, such as Biology (e.g., neuroscience networks), Chemistry (e.g., protein interaction networks), Physics, Information Technology (e.g., WWW, Internet), Economics (e.g. interbank payments) and Social Sciences (e.g. collaboration and acquaintance networks)~\cite{barabasi2016network}. Despite the fact that networks can originate from different domains, most of them share a few common characteristics such as scale-free and small-world properties~\cite{barabasi1999emergence,watts1998}, highly clustered nature~\cite{newmancluster} and sparseness~\cite{sparse}. 

Modeling real networks is of great interest, since it may help to understand the underlying mechanisms and principles governing the evolution of networks. Moreover, such models are mathematically tractable and allow for rigorous analysis. Furthermore, an appropriate model preserves the key characteristics, yet ensures the anonymity of the original real network~\cite{fong2009privacy}. 

Throughout the years, several network models have been proposed to gain a better understanding of real-world networks, for an extensive overview see the paper of Goldenberg \textit{et al.}~\shortcite{goldenberg2010survey}. However, without attempting to be comprehensive, the most influential models include the scale-free preferential attachment model introduced by Barabási \& Albert~\shortcite{barabasi1999emergence}, the small-world model proposed by Watts \& Strogatz~\shortcite{watts1998}, and the highly clustered community structure model of Newman~\shortcite{newmancluster}, each of them was motivated by some of the aforementioned common characteristics of real networks. 

In order to further characterize the structure of networks, numerous graph metrics have been introduced in the past two decades such as degree-distribution-related measures~\cite{aliakbary2014quantification} and other node-level centrality measures, local and global clustering coefficients, assortativity coefficient, density~\cite{barabasi2016network}, moreover motif related attributes~\cite{milo2002network, prvzulj2007biological}.  Naturally, there is significant redundancy among these measures and there is a great deal of effort in studying the correlation between the metrics and identifying a non-redundant selection of measurements that describes every aspect of networks~\cite{bounova2012overview,garcia2013correlation,jamakovic2008relationships,marcell2018data}. 

\begin{figure}
    \centering
    \includegraphics[width=\textwidth]{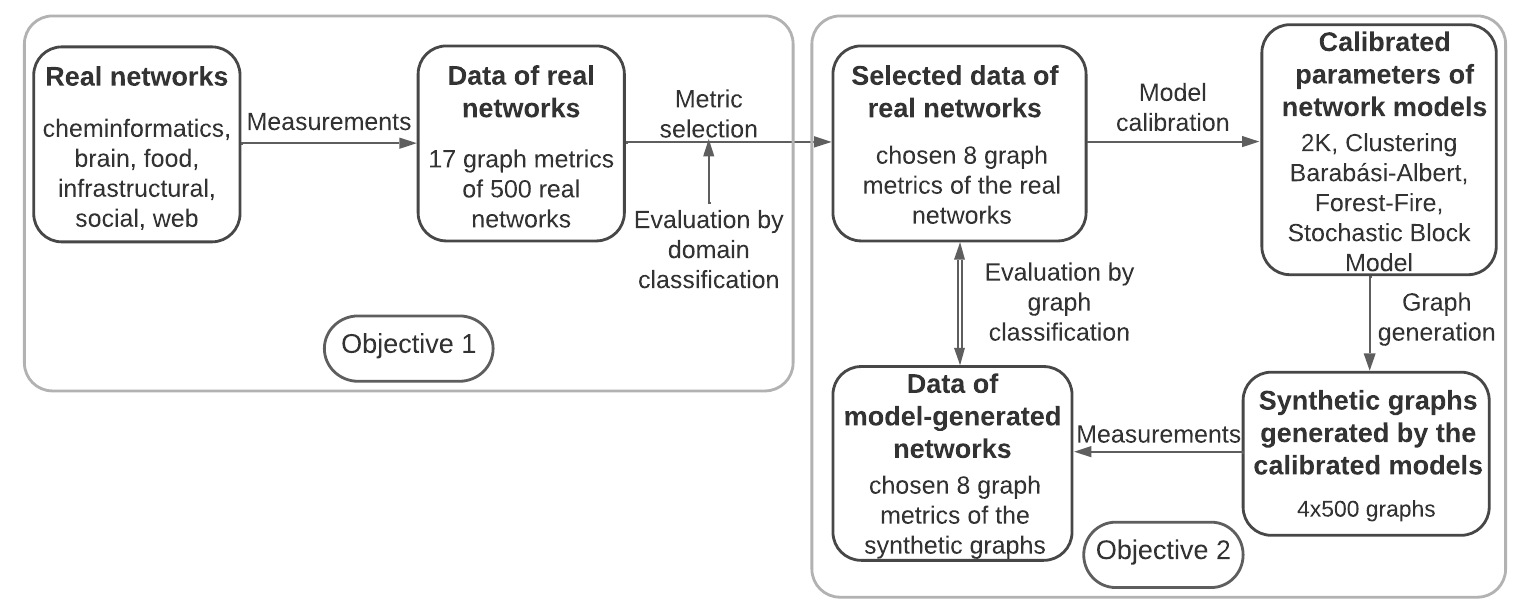}
    \caption{The workflow of the proposed methodology.}
    \label{fig:workflow}
\end{figure}

The present research  was driven by the following two main objectives (see Figure~\ref{fig:workflow}):
\begin{enumerate}
    \item Studying the pair-wise correlations and descriptive power of structural graph metrics of real networks to find a minimal set of metrics that describes networks as accurately as possible. \label{obj_1}
    \item Understanding how well measurement-calibrated network models can describe real networks from different domains and investigating what structural properties the models can and cannot capture. \label{obj_2}
\end{enumerate}

The outline of this paper is as follows: in Section \ref{related_works} we give a thorough overview of related literature categorized by topics. The diverse scope of the literature review is due to the unifying nature of the present paper that joins different branches and building blocks of data-driven analysis of complex networks. In Section \ref{data_and_methods} we describe our selection of real networks, the investigated graph metrics, and the proposed network model calibration procedures. Furthermore, we show that despite the randomness of these models, their structural properties are stable enough to perform \textit{grid optimization} based parameter calibration. In Section \ref{analysis_of_network_domains}  we investigate the structural properties of real networks by studying the correlation profiles of graph measurements across network domains and compose a minimal set of metrics that has high descriptive power and low redundancy that we use later in this work (objective \ref{obj_1}). In Section \ref{analysis_of_network_models}, we propose a machine-learning-based framework for calibration and evaluation of the network models using a given set of graph features (objective \ref{obj_2}). Particularly, we examine how realistic graphs a network model can generate, which domains are easy to model, and what structural characteristics of real networks are difficult to capture with network models. Section \ref{conclusion} concludes the work.

\section{Related work \label{related_works}}
Our paper relates to recent studies that reflect the growing interest in the identification of the most distinguishing structural network features and assessing structural similarity between networks. Our literature review is organized according to the following questions:
\begin{enumerate}
    \item What structural properties should be considered in model calibration, i.e., what are the most distinguishing features of networks?
    \item How to measure the similarity between networks, in particular between model-generated graphs and real networks?
    \item What machine learning techniques should be used to distinguish between various classes of networks, in particular between model-generated graphs and real networks?
    \item What time-efficient optimization or statistical approximation method should be used for model parameter calibration?

\end{enumerate}

There are no canonical methods to answer the above questions, several approaches have been proposed in the literature throughout the years, which we summarize in the following subsections. Furthermore, we also highlight how this work contributes to the literature.  

\subsection{Graph metric selection}
There is a great deal of effort related to finding a non-redundant size-independent set of graph measurements. For example, Garcia-Robledo \textit{et al.}~\shortcite{garcia2013correlation} follow a data-driven approach to study the Pearson correlation of different properties of evolving Internet networks and apply clustering techniques to find a non-redundant set of metrics. 

Bounova \& de Weck~\shortcite{bounova2012overview} give a great overview of network topology metrics and their normalization, they also carry out correlation and principal component analysis on a relatively large dataset that consists of both model-generated graphs and real networks. Similarly, Filkov \textit{et al.}~\shortcite{filkov2009modeling} investigate both network models and a collection of 113 real networks to find a set of metrics, which enables comprehensive comparison between any two networks. Furthermore, Jamakovic \& Uhlig~\shortcite{jamakovic2008relationships}  study the correlation of metrics and visual comparison of distribution node-level features using 13 real networks. 

Sun \& Wandelt~\shortcite{sun2014network} use a dataset of air navigation route systems and investigate functional dependencies among graph metrics through regression analysis to select a less redundant set of metrics. Their metric selection relies on a graph, where the nodes correspond to the metrics and an edge is drawn between two nodes if the predictive power (measured in $R^2$) of the corresponding metrics have on each other is above a certain threshold. A metric is selected from each connected component of this auxiliary graph. Harrison~\shortcite{harrison2014network} gives an in-depth study of the behavior of network metrics of model-generated graphs. Moreover using statistical learning techniques (Fisher's method and ROC curve analysis) the author also investigates graph similarities through the distinguishing power of different selections of metrics.

Relying on the reviewed literature, we also propose a similar metric selection procedure. However, in contrast to the majority of the aforecited papers, in this paper we use Spearman's rank correlation coefficient instead of the Pearson correlation since rank correlation can measure non-linear relationships as well, moreover, it is less sensitive to outliers~\cite{croux2008robustness}, hence it provides us more useful information about the relationship of the metrics. 

\subsection{Graph similarity \label{graph_sim}}
Another related area of research is concerned  with  estimating the similarity between graphs with unknown node correspondences. Similarities can be defined in numerous ways, for instance, based on variations of \textit{graph isomorphism}, which involves subgraph and degree isomorphism~\cite{kelmans1976comparison, wilson2008study} and edit distances~\cite{gao2010survey}. Another approach utilizes \textit{graph kernels}, which are kernel functions that compute inner products on graphs measuring their similarity~\cite{ralaivola2005graph,vishwanathan2010graph, kashima2004kernels, kang2012fast, hammond2013graph}. \textit{Motif counting} is an alternative frequently used technique~\cite{milo2002network, prvzulj2007biological, ikehara2017characterizing, faust2006comparing, bordino2008mining, janssen2012model}, which refers to the investigation of the distribution of small \say{graphlets} (e.g., all graphs of size 3 or 4). NetEmd, a network comparison method introduced by Wegner \textit{et al.} \shortcite{wegner2018identifying} is also based on the distribution of motifs.  \textit{Network Portraits} provide an additional method to compare complex networks as well as to visualize their structural characteristics~\cite{bagrow2008portraits}. 

A significantly different technique applied by Leskovec \textit{et al.}~\shortcite{leskovec2010kronecker} and Sukrit \textit{et al.}~\shortcite{sukrit2016kronfit}, which works by first fitting the Kronecker Graph model to the networks, and then comparing the fitted parameter matrices. Furthermore, a similarity metric can also be defined with the help of $dK$-distributions~\cite{sala2010measurement, mahadevan2006systematic}. A drawback of the latter two methods is that they are computationally highly expensive.

In a recent work, Mheich \textit{et al.}~\shortcite{mheich2018siminet} proposes an alternative algorithm called \textit{SimiNet} for measuring similarity between graphs, which takes into account the physical location of the nodes, i.e. \textit{SimiNet} requires prior knowledge of the coordinates of the nodes lying in a 3D coordinate system. Schieber \textit{et al.}~\shortcite{schieber2017quantification} introduce a novel measure for network comparison, which is based on the Jensen--Shannon differences between the sets of node-distance probability distributions defined for each node. Chen \textit{et al.}~\shortcite{chen2018complex} propose another method based on the Jensen--Shannon divergence, referred to as \textit{communicability sequence entropy} that considers the communicability between nodes which is related to the path lengths.


In this work, we apply another possible calculation of network similarity, which is based on a feature vector representation of graphs. These feature vectors can contain global information of the graphs (e.g., average degree, diameter, and z-score of a motif) or can be obtained by graph embedding methods~\cite{rossi2017deep, narayanan2017graph2vec}. The derived vectors from the graphs are usually compared with the traditional distance functions, such as Canberra~\cite{bonner2016efficient}, weighted versions of Manhattan and Euclidean distances~\cite{aliakbary2015noise, sun2014network} or a suitable distance function can be constructed with \say{Distance Metric Learning}~\cite{ktena2017distance, aliakbary2015distance, yang2006distance,weinberger2009distance}.

By the supervised nature of distance (or similarity) metric learning, it requires pairs of graphs with known distances (or similarities), which are usually unknown in reality. The usual approach to overcome this problem is to consider two graphs similar if they both originate from the same domain or class and dissimilar otherwise. For example, this method is well-suited if one aims to  distinguish between  certain types of networks, e.g., normal vs.~cancer cell networks, or connectome networks of patients with and without autism spectrum disorder~\cite{ktena2017distance}. On the other hand, this technique is not applicable in our case, since it is one of our research goals to reveal how similar the network domains are without any prior assumptions about their similarity.

In a recent work, Bai \textit{et al.}~\shortcite{bai2019simgnn} propose an alternative method to calculate graph similarity with efficient time complexity. Their machine-learning-based approach combines two strategies, namely, it performs both graph-level and node-level embedding to obtain fine-grained information of the graphs. The final similarity score is calculated through a deep neural network whose inputs are the results of the embeddings. In order to train the deep neural network, the authors used a ground-truth similarity score based on the Graph Edit Distance (GED) of the training graphs~\cite{gao2010survey}. This also can be seen as a disadvantage, since the method actually approximates the GED of two networks. However, as Schieber \textit{et al}~\shortcite{schieber2017quantification} showed with a simple example, GED is not a perfect graph distance metric, since graphs that are close to each other according to GED can have totally different structural properties. 

Mart\'inez \& Chavez~\shortcite{martinez2019comparing} have recently shown that Euclidean distance (defined by the Frobenius norm of the difference of adjacency matrices) of equal-sized graphs can be more effective (or at least provide a better trade-off between accurate and fast performance) than other more complicated methods introduced by Hammond \textit{et al.}~\shortcite{hammond2013graph} and Schieber \textit{et al.}~\shortcite{schieber2017quantification}.

Finally, Soundarajan \textit{et al.}~\shortcite{soundarajan2014guide} give a great overview, categorization, and guidance for network similarity methods reflecting the state-of-the-art techniques in 2014.

In this paper, we investigate the similarity of networks stemming from different domains and models based on the Canberra distance, because according to Bonner \textit{et al.}~\shortcite{bonner2016efficient}, it is the most suitable distance metric to compare graphs through their feature vectors.

\subsection{Network classification}
Using classification techniques to distinguish between real and model-generated graphs has attracted a lot of research interest recently. However, the perspective of these studies differs from ours. The vast majority of related papers do not calibrate the parameters of the models, but rather focus on the \textit{spanning} graph space of the models, i.e. they consider a network model as a special domain of networks. 
Naturally, with certain parameter settings, every model can generate graphs that are not similar to real graphs at all, which can bias the distinguishing power of graph metrics. Hence, that is why in this work we are focusing on the parameters of the models that induce as realistic graphs as the given model is capable of.   

Janssen \textit{et al.}~\shortcite{janssen2012model} propose a method to choose a network model that best fits a given real network. Their model selection consists of three steps: first, they generate 6,000 graphs with six well-known models and extract graph metrics and graphlet distributions from these graphs. The parameters of the models are randomly sampled such that the size and the density of the generated graphs are similar to the target networks'. Second, they train an Alternating Decision Tree classifier on the previously obtained dataset, where the target variable is the type of the model. Finally, they compute the same feature vectors for Facebook networks and then feed these vectors to the trained classifier to see how well the models fit the given networks.  

Ikehara \& Clauset~\shortcite{ikehara2017characterizing} apply machine learning tools to investigate the clustering coefficient, the distribution of motifs and degrees in 986 real and 575 model-generated graphs in order to find the most distinguishing features of real and synthetic networks. In this work, we do not consider the distribution of motifs because these attributes are rather difficult to interpret, furthermore, their calculation is computationally expensive.

Canning \textit{et al.}~\shortcite{canning2018predicting} use Random Forest to predict the origin (network domains and type of the models) of 402 real and 383 generated graphs, however, they use metrics such as the total number of triangles and density, which are strongly correlated with the size of the network, which is a very strong distinguishing feature in network classification.

Bonner \textit{et al.}~\shortcite{bonner2016deep} propose a Deep Learning based approach, called \textit{Deep Topology Classification}, to classify graphs by both global and local features. They evaluate their methodology on a large dataset consisting of synthetic graphs generated by five well-known network models with heuristically chosen parameters. 

Barnett \textit{et al.}~\shortcite{barnett2019endnote} recommend using the random forest algorithms to solve network classification tasks based on manually selected network features. We will follow a similar procedure in this work.

While most of the related works focusing on graph classification investigate a particular network domain such as protein interaction networks~\cite{middendorf2005inferring}, chemical compounds and cell-graphs (brain, breast, and bone)~\cite{li2012effective, ralaivola2005graph, ktena2017distance} or Facebook networks~\cite{sala2010measurement, ugander2013subgraph, janssen2012model}, in this paper we study six different network domains and their model-generated counterparts.

\subsection{Model calibration}
A closely related work is by Sala \textit{et al.}~\shortcite{sala2010measurement} who fit six network models to four large Facebook networks by grid search parameter calibration. For the parameter fitting, they use a $dK$-series based similarity metric~\cite{mahadevan2006systematic} and evaluate the \say{fidelity} of the fitted models by calculating the Euclidean distance between graph measurements derived from the target and the model-generated graphs. 

Another similar paper is by Bl\"asius \textit{et al.}~\shortcite{blasius2018towards} who fit four random graph models to 219 real-world networks (mostly from Facebook) and evaluate their goodness-of-fit with respect to different selections of graphs measurements by investigating the failure rates of a Support Vector Machine classifier that predicts whether the corresponding vector of graph metrics belongs to real or model-generated networks. 

Aliakbary \textit{et al.}~\shortcite{aliakbary2015noise} follow a two-step approach to first find the most suitable model for a given real network and secondly to fine-tune its parameters. In their work, first, they generate graphs with different models using randomly chosen parameter settings to obtain the spanning graph space of each model. Then using a distance learning method, a distance metric function is defined that distinguishes well networks that were generated by different models. They use a Nearest Neighbour (kNN) classifier with the previously defined distance function to assign a model to each target real network. The parameters of the previously assigned model are fine-tuned by Artificial Neural Networks (ANN). 

In a follow-up paper, Attar \& Aliakbary~\shortcite{attar2017classification} extend the previous network calibration framework as follows: instead of generating graphs with random parameter settings as an input for the distance learning method, the authors first fit each network model to the target networks (using ANN). To train the distance learning, they generate a few graph instances for each target network with the fitted parameters of each model, and finally, the best performing model is selected with the former kNN classifier. 
A limitation is that neural networks usually require large datasets to perform well, furthermore as Bl\"asius \textit{et al.}~\shortcite{blasius2018towards} also pointed out, this method in its original form is not able to identify the structural characteristics of real networks that models can or cannot capture (cf. objective~\ref{obj_2}).

While the afore-cited papers apply search optimization techniques, in some cases statistical parameter estimating is also possible. In the following works, a network model is considered as a probability distribution over graphs. Bezáková \textit{et al.}~\shortcite{bezakova2006graph} propose a Maximum-Likelihood-based parameter estimation (and comparison) method for four graph models. Recently, Arnold \textit{et al.}~\shortcite{arnold2021likelihood} introduced a likelihood-based approach to find which growing model is more likely responsible for the generation of a network. Their work also involves multiple related tasks such as parameter calibration, graph similarity, and network classification. 
Fay \textit{et al.}~\shortcite{fay2014graph} study Approximate Bayes Computation as a technique for estimating the parameters of network models relatively to an observed graph. An additional brief survey of generating realistic synthetic graphs is given by Lim \textit{et al.}~\shortcite{lim2015survey}.

 In this work, we also apply a sophisticated model calibration technique (see Section \ref{model_fitting}) on 500 networks from six different domains.

\section{Data and Methodology \label{data_and_methods}}

 This study relies on our large dataset: we calculated 17 metrics of 500 real-world networks and a subset consisting of 8 reasonably chosen (see Section \ref{real_networks}) metrics for $4\times500$ synthetic graphs by generating four counterparts for each real network with well-known network models. The dataset of measurements is available online~\cite{supplementary}. In this section, we detail the composition of our dataset and describe the main challenges and our method of model calibration.

\subsection{Data of real networks \label{real_networks}}
Networks from six different domains are studied in this paper, namely food, social, cheminformatics, brain, web, and infrastructural networks. The collected networks are also available online~\cite{supplementary}.
Table~\ref{tableofnetworks} briefly describes the composition of network domains.

\begin{table}[]
\caption{Composition of the collected set of real networks.}
\label{tableofnetworks}
\scalebox{0.85}{
\begin{tabular}{llcc}
\textbf{Domain} & \textbf{Description} & \multicolumn{1}{l}{\textbf{Range of network sizes}} & \multicolumn{1}{l}{\textbf{Number of networks}} \\ \hline
Brain & Human and animal connectomes & 50-2,995 (avg: 946) & 100 \\
Cheminformatics & Protein-protein (enzyme) interaction networks & 44-125 (avg: 55) & 100 \\
Food & What-eats-what, consumer-resource networks & 19-1,500 (avg: 117) & 100 \\
\multirow{2}{*}{Infrastructural} & \multirow{2}{*}{\begin{tabular}[c]{@{}l@{}}Transportation (metro, bus, road, airline)\\ and distribution networks (power and water)\end{tabular}} & \multirow{2}{*}{39-40K (avg: 4,562)} & \multirow{2}{*}{68} \\
 &  &  &  \\
Social & Facebook, Twitter and collaboration networks & 86-34K (avg: 6,183) & 118 \\
Web & Samples of the World Wide Web & 146-16K (avg: 4,488) & 14
\end{tabular}}
\end{table}

The graphs were gathered one by one from online databases, namely Network Repository~\cite{networkrepository}, Index of Complex Networks~\cite{icon}, NeuroData's Graph Database~\cite{neurodata}, The Koblenz Network Collection~\cite{konect}, Interaction Web Database~\cite{iwdb}, Transportation Networks for Research~\cite{transportnetworks} and Centre for Water Systems \cite{water_dist_networks}. After importing the graphs, we removed self-loops and treated them as undirected, unweighted graphs. 

\begin{figure}[h]
    \centering
    \includegraphics[width=0.3285\linewidth]{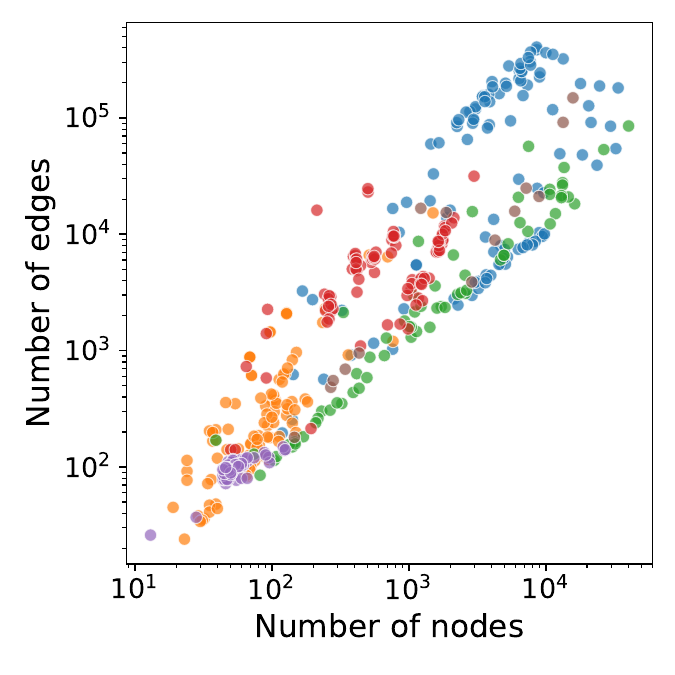}
    \includegraphics[width=0.3285\linewidth]{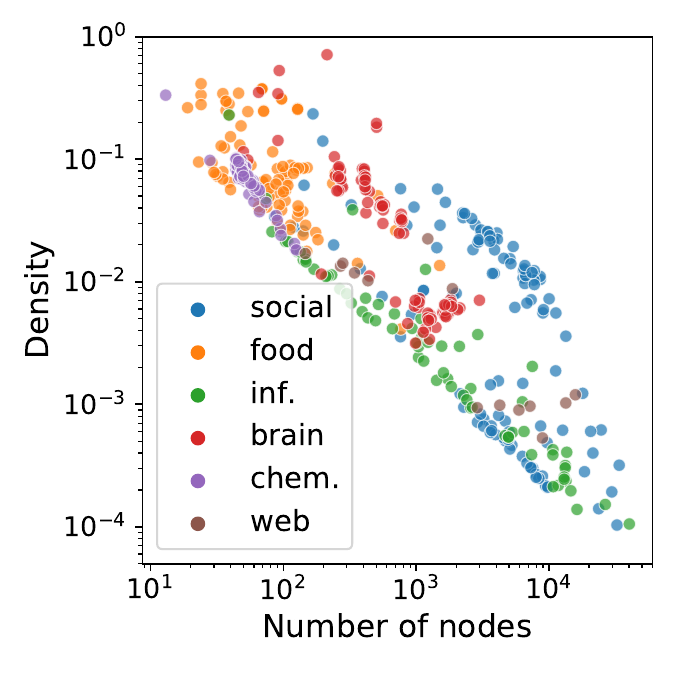}
    \includegraphics[width=0.3285\linewidth]{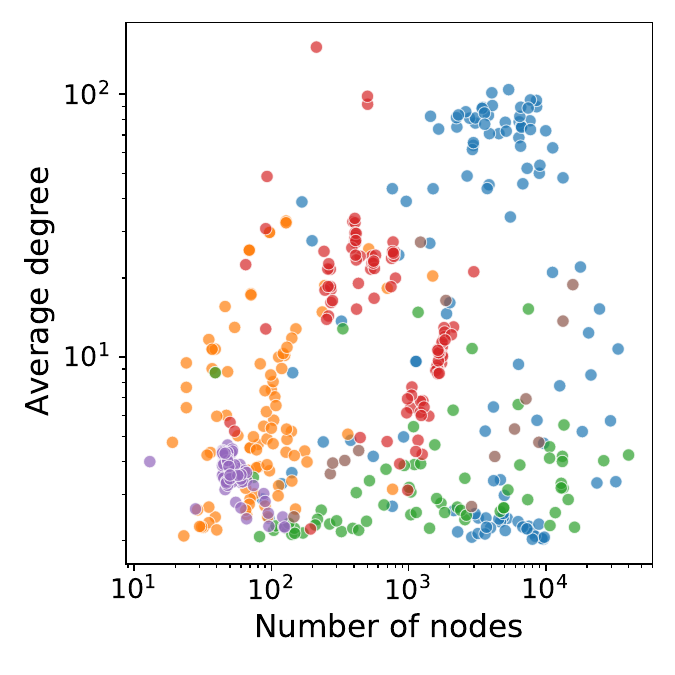}
    \caption{The scatter plot of the real networks. The points are colored according to the network domain. On the left: the log-log plot of the number of edges against the number of nodes. The center figure shows the the log-log plot of the graph density versus the number of nodes. On the right: the log-log plot of the average degree against the number of nodes.}
    \label{fig:size_density_avgdeg}
\end{figure}

The left side of Figure~\ref{fig:size_density_avgdeg} shows the range of the number of nodes and edges of the collected networks. In Figure~\ref{fig:size_density_avgdeg}, the slope (gradient) of the left scatter plot illustrates the sparseness of real networks, i.e., the number of edges grows rather linearly than quadratically. Furthermore, in the center scatter plot the slope is roughly minus one which means that the density decays hyperbolically with the number of nodes, i.e., $density \sim 1/size$. Comparing the center and the right figure of Figure~\ref{fig:size_density_avgdeg}, we can clearly observe that the average degree is less size-dependent than the graph density.

\subsection{Metrics and notations \label{metrics_and_notations}}
First, we calculate the following rich set of metrics for the real networks: assortativity, average clustering coefficient, average degree, average path length, density, global clustering coefficient, four interval degree probabilities~\cite{aliakbary2014quantification}, largest eigenvector centrality, maximum degree, maximum edge and vertex betweenness centralities, number of edges and nodes and pseudo diameter. If one aims to investigate these measurements for significantly different-sized networks, some normalizations are required, in particular in the case of the average path length, diameter, and maximum degree. Relying on Bounova \& de Weck~\shortcite{bounova2012overview}, we normalized the maximum degree with the number of nodes minus one, however, as opposed to Bounova \& de Weck, instead of the longest possible path length, we normalized the average path length and the diameter with the logarithm of the size of the graph to obtain less correlated variables with the network size. The reason why the average degree does not require normalization is that its normalized version (with the size minus one) is the graph density~\cite{bounova2012overview}. The graph density $D$ is defined as the ratio of the number of edges $|E|$ with respect to the maximum possible edges, i.e., $D = |E| / \binom{|V|}{2} = \frac{|E|}{|V| (|V|-1)}$.

In order to quantify the degree distribution of the networks, we apply the \textit{interval degree probability} method introduced by Aliakbary \textit{et al.}~\shortcite{aliakbary2014quantification}, which is the computation of histogram values with special different sized bins that are defined using the mean and the standard deviation of the degrees. In contrast to the work of Aliakbary \textit{et al.}~\shortcite{aliakbary2014quantification} we defined four bins instead of eight because we found that otherwise many of the bin intervals end up empty. Hence, we quantify the degree distribution of the graphs with four values of interval degree probabilities. Formally, the interval degree probabilities (IDP) can be defined as follows: 

For a given graph $G$, let $\mu$ and $\sigma$ be the average and the standard deviation of the degree distribution of $G$. Then we define the following four intervals $R(i)$, $i=1,\ldots,4$:
\begin{equation}
    R(i) = 
    \begin{cases}
        [\,1, \,\,\mu - \sigma\,), & \text{if } i=1\\
        [\,\mu - \sigma,\,\, \mu\,), & \text{if } i=2\\
        [\,\mu,\,\, \mu + \sigma\,), & \text{if } i=3\\
        [\,\mu + \sigma,\,\, \infty\,), & \text{if } i=4
    \end{cases}
\end{equation}

The interval degree probabilities $IDP(i)$, $i=1,\ldots,4$ are defined as follows:
$$
IDP(i) = P(\deg(v)\in R(i)), \quad v\in V(G).
$$

\subsection{Correlation network}
Throughout this paper, we often refer to the 
\textit{correlation network}, which is frequently used object in the literature~\cite{garcia2013correlation, jamakovic2008relationships, sun2014network, blasius2018towards}. A correlation network is an undirected weighted graph with nodes corresponding to variables (in our case graph metrics) and two nodes are connected if their absolute correlation (in our case Spearman's rank correlation) is greater than a given threshold. The weights of the links are the values of the absolute correlations.

In this work, the application of the correlation network is three-folded: we use it to visualize the strong universal correlations of graph metrics, it is also a key component of the metric selection, furthermore, we propose a novel technique for  quantifying a similarity score between groups of networks, based on the adjacency matrices of the correlation networks corresponding to different groups (e.g., domains and types of network models).

\subsection{Model calibration \label{model_fitting}}
For each real network, we generated additional four graphs by four different models with previously calibrated parameters to mimic real networks as accurately as possible. We study  how  easy  it  is  to  calibrate  the  network  models  and  what descriptive ability they have with respect to real-world networks. The four chosen models are the clustering version of the Barab\'asi--Albert model~\cite{holme2002growing}, the 2K-Simple model~\cite{gjoka2015construction}, which captures the joint-degree distribution of a graph, the forest-fire model~\cite{leskovec2005graphs}, which was motivated by citation patterns of scientists, and the (degree-corrected microcanonical) stochastic block model~\cite{peixoto2017nonparametric} that creates graphs with community structure. The 2K and the stochastic block models are more complex models than the other two and can be considered state-of-the-art models for synthesizing real-world networks while preserving their key structural properties.

With a given target network $G_T$ (real network), a network model $M(\theta)$ with parameter vector $\theta$ and a similarity $s$ (or distance $d$) function defined on graphs pairs, the goal of model calibration is to find a parameter vector $\theta^*$ of the model, such that the generated graph $G_{M(\theta^*)}$ (realization of $M(\theta^*)$) is as similar (or as close) to the target network as possible. Formally it can be expressed as:
\begin{equation}
\label{eq:calibration}
    \theta^* = \argmax_\theta \,s(G_{M(\theta)},\, G_T) =  \argmin_\theta \, d
(G_{M(\theta)},\, G_T). 
\end{equation}

Note that one can maximize the average similarity (or minimize the distance) between the target network and a few realizations of the model with the same parameters. That is:
\begin{equation}
     \theta^* = \argmax_\theta\, \frac{1}{n}\sum_{i=1}^n s\left(G^{(i)}_{M(\theta)},\, G_T\right) =  \argmin_\theta \, \frac{1}{n}\sum_{i=1}^n d
\left(G^{(i)}_{M(\theta)},\, G_T\right), 
\end{equation}
where $n$ is the sample size of the independent identically generated (i.e. generated with the same $\theta$ parameter vector) graphs $G^{(1)}_{M(\theta)}, \ldots, G^{(n)}_{M(\theta)} $. 
However, in this work we only consider one realization of the models for two reasons, first, we show in the following subsection that the models are stable enough to perform parameter calibration as defined in \eqref{eq:calibration}, furthermore, the calculation of average similarity (or distance) would be more computationally expensive. 

In this work, the $d$ distance function is defined as the Canberra distance of a reasonably chosen non-redundant selection of graph metrics, formally: $d(G_1,G_2) = d_{\text{Can}}(f(G_1),f(G_2)),$ where $f = (f_1, f_2, \ldots, f_k)$, and $f_i$'s are real-valued functions defined on graphs (i.e., graph metrics) and $k$ is the number of selected metrics. The process of metric selection is detailed later in Section \ref{analysis_of_network_domains}. The Canberra distance of two $k$-dimensional vectors is defined as $d_{\text{Can}}(x, y) = \sum_{i=1}^k\frac{| x_i - y_i|}{|x_i| + |y_i|}$.

The distance minimization in this paper is carried out with grid search, particularly for a given $M(\theta)$ model, where $\theta = (\theta_1, \theta_2, \ldots, \theta_p)$, we perform a grid search on the $p$ dimensional state space of the parameters. In what follows we briefly outline the parameter calibration of each network model that we investigate in this work.

\subsubsection{Clustering Barab\'asi--Albert model}
The Clustering Barab\'asi--Albert (CBA) model, introduced by Holme \& Kim~\shortcite{holme2002growing}, has three input parameters: $n$ number of nodes, $m$ new edges of a newcomer vertex, and $p$ probability of triad formation. This $p$ parameter influences the clustering characteristic of the generated network. Holme \& Kim~\shortcite{holme2002growing} showed that for fixed $n$ and $m$, there is a linear relationship between $p$ and the average clustering coefficient. The parameter calibration of the CBA model is as follows: Let $G_T$ denote the target network, furthermore $\bar{d}$ denotes the average degree of $G_T$. Then the parameter $n$ of CBA is simply the size of $G_T$, and $m$ and $p$ are chosen by:
$$(m, p) = \argmin_{(k,q)\in M \times P} d\big(G_{\text{CBA}(n,k,q)}, G_T\big),$$
where $M = \left\{\round{\bar{d}/2-1},\, \round{\bar{d}/2},\, \round{\bar{d}/2+1 }\right\}$, $P = \{0,0.05,0.1,\ldots,1 \}$, $M\times P$ is the Cartesian product of $M$ and $P$ and $\round{\text{ . }}$ denotes the rounding (nearest integer) function. The reason behind the choice of set $M$ is that CBA model with $m = \round{\bar{d}/2}$ parameter setting produces graphs with $e = n\cdot \round{\bar{d}/2}$ edges, i.e. the average degree in the generated graphs is $2e/n$ which is around $\bar{d}$ as one would desire. We also allow for a small deviation from this value in the minimization that may lead to a better fit overall. For example, it may happen, that if the value of the parameter $m$ is slightly smaller (or larger) than the half of the average degree of the target graph, then it leads to a better fit regarding the other graph measurements of the generated graphs. For generating clustering CBA graphs we applied the \texttt{powerlaw\_cluster\_graph} function of NetworkX~\cite{hagberg2008exploring}.

\subsubsection{Forest-fire model}
The forest-fire model (FF) introduced by Leskovec \textit{et al.}~\shortcite{leskovec2005graphs} has two parameters: $n$ number of nodes and $p$ burning probability. Using the previous notations, $n$ is set to be the size of $G_T$ target graph and we minimize in parameter $p$ as follows:
$$p = \argmin_{q\in Q} d\big(G_{\text{FF}(n,q)}, G_T \big),$$
where $Q=\{0.05,0.1,0.15,\ldots,0.90, 0.95\}$. The reason why $q$ is between $0.05$ and $0.95$ in the minimization, is that $q$ must be positive and less than $1$, moreover, the structure of the generated graph is not sensitive to $q$ if its value is smaller than $0.05$ or greater than $0.95$. In our analyses we used the \texttt{Forest\_Fire} function of python-igraph~\cite{csardi2006igraph}.

\subsubsection{2K model}
To fit the 2K model~\cite{gjoka2015construction}, only the joint degree distribution of the target network is needed. We used the \texttt{joint\_degree\_graph} function of the NetworkX~\cite{hagberg2008exploring} module to generate simple random graphs with the given joint degree distribution. Let $k_{j,l}$ denote the number of edges in the target graph whose endpoints' degree is $j$ and $l$. Then the 2K model builds the graph, such that it copies the nodes of the target graph and then in each iteration, the algorithm picks two disconnected nodes $v$ and $w$ of degree $j$ and $l$ correspondingly, for which the $k_{j,l}$ in the random graph has not reached its target yet, and it adds the edge $(v, w)$ to the synthetic graph.

\subsubsection{Stochastic block model}
The degree-corrected stochastic block model can be fitted by the \texttt{minimize\_blockmodel\_dl}, and random graphs can be generated using the \texttt{generate\_sbm} functions of the graph-tool package~\cite{peixoto_graph-tool_2014}. We applied the microcanonical version of the model~\cite{peixoto2017nonparametric}. The stochastic block model is fitted by minimizing its description length using an agglomerative heuristic. 

\subsection{Stability of the models}
Since the network models generate random graphs, the question naturally comes up: How robust are the graph metrics of the fitted models with fixed parameters?
We have analyzed the sensitivity of the models on six different-sized graphs from different domains, the chosen networks are detailed in Table \ref{sixgraphs}.  For each of these six real networks, we fitted each network model and then generated 30 graph instances with each model using the previously fitted parameter settings. For the sensitivity analysis, we studied the distribution of the graph measurements of the graph instances.

\begin{table}[h]
\centering
\begin{tabular}{llrr}
\multicolumn{1}{c}{\textbf{Domain}} & \multicolumn{1}{c}{\textbf{Name}} & \multicolumn{1}{c}{\textbf{Size}} & \multicolumn{1}{c}{\textbf{\makecell[c]{Number of\\ edges}}} \\ \hline
Social & \href{https://snap.stanford.edu/data/ca-AstroPh.html}{ca-AstroPh}~\cite{leskovec2007graph} & 17,903 & 196,972 \\
Web & \href{http://santafe.edu/~aaronc/shared/}{Darkweb}~\cite{griffith2017graph} & 7,178 & 24,879 \\
Brain & \href{http://awesome.cs.jhu.edu/graph-services/download/}{Jung2015}~\cite{kiar2016gremlin} & 2,989 &  31,548\\
Infrastructure & \href{https://github.com/achatterjee3/Dataset}{ABN}~\cite{chatterjee2016statistical} & 1,113 & 2,150 \\
Food & \href{https://www.nature.com/articles/srep21179#supplementary-information}{Srep}~\cite{dunne2016roles} & 235 & 1,743  \\
Cheminformatics & \href{http://networkrepository.com/ENZYMES-g292.php}{Enzymes-g292}~\cite{canning2018predicting}  & 60 & 100 

\end{tabular}
\caption{Description of chosen graphs for stability analysis. The names of the graphs are also hyperlinks to their sources, moreover, these graphs are also analyzed in the cited papers.}
\label{sixgraphs}
\end{table}

\begin{figure}[h]
    \centering
    \includegraphics[width=\textwidth]{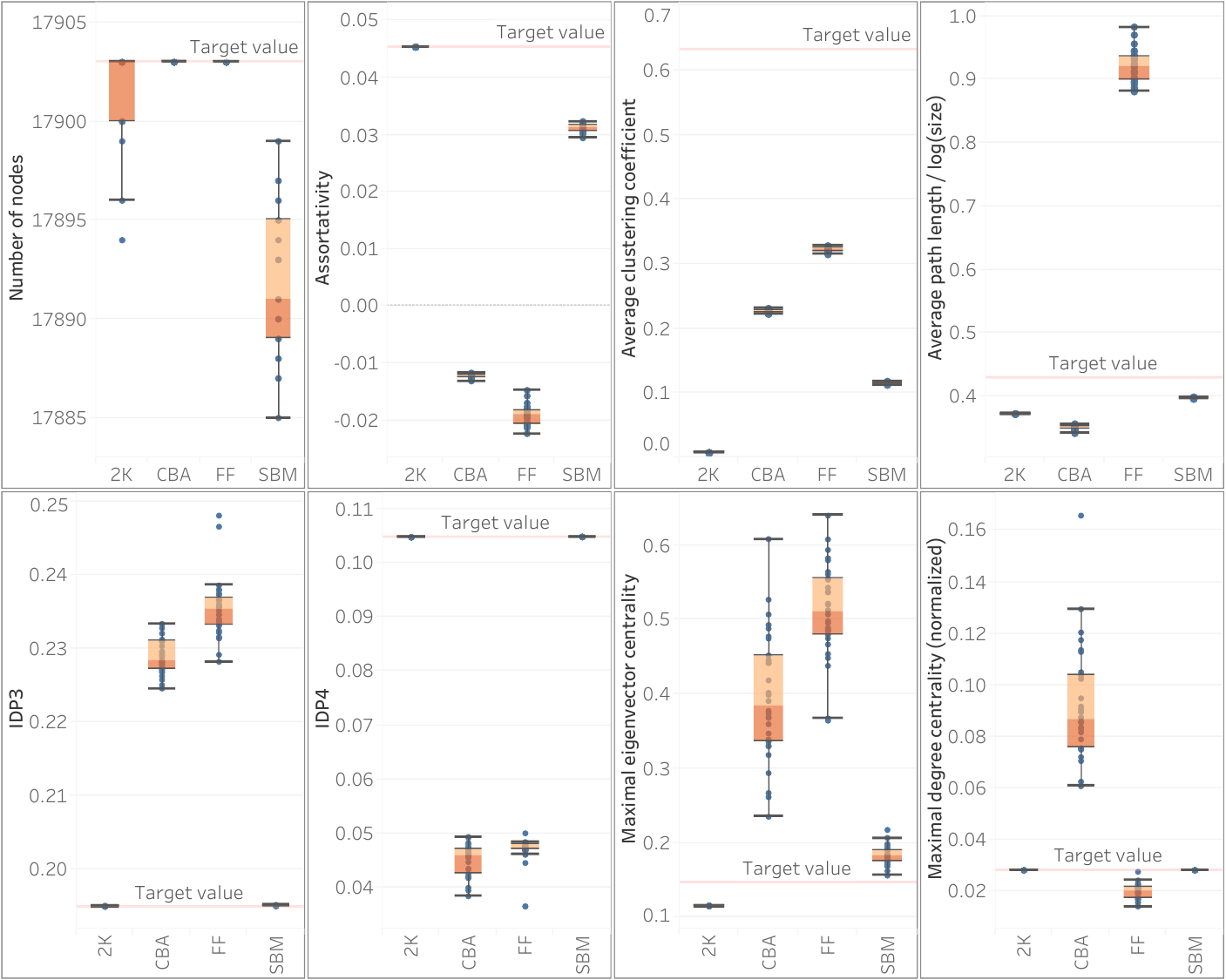}
    \caption{Boxplots of the metrics of the measurement-calibrated models on a social network, namely on the the Astro Physics collaboration network.}
    \label{fig:boxplotsSocial}
\end{figure}

Figure \ref{fig:boxplotsSocial} shows the distribution of the metrics of the measurement-calibrated models for the social network that has 17,903 nodes. Since the 2K model captures all the information about the joint degree distribution, when it generates a connected graph, it is capable of exactly capturing the assortativity, the interval degree probabilities, and the degree centrality. Overall, the models could not estimate the average clustering coefficient well, not even the clustering Barabási--Albert model. Besides the clustering coefficient, the models also have difficulties capturing the average path length and the eigenvector centrality. Generally, the SBM and 2K models are more robust and \say{stable}, than the CBA and FF models.

The boxplots for the other domains can be found in the supplementary material \cite{supplementary}. The results suggest that as the size of the networks increases, the variance of the graph measurements of the fitted models decreases, however, the variance of the metrics is already relatively small on the smaller graphs. 

We can conclude, that the characteristics of the calibrated models are stable enough to take a single generated graph as a representative, and it makes it reasonable to analyze the goodness-of-fit of the network models and perform machine learning tasks.

Note that reason why the number of nodes is not fixed for the 2K and the SBM models is that these models do not always generate connected graphs, and in this study we took the largest connected components of both the real-world and the model-generated networks. Furthermore, the reason why IDP1 and IDP2 are not shown in Figure \ref{fig:boxplotsSocial}, is that first, it is enough to study three of these metrics, since the sum of them gives one. Moreover, in this work, we selected IDP1, IDP3, and IDP4 (the selection process is detailed later in Section \ref{correlation_of_metrics}). The reason why IDP1 is not shown in Figure \ref{fig:boxplotsSocial}, is that for this specific network (and its model-generated counterparts), the value of IDP1 is zero in all cases (due to the heavy-tailed degree distribution).

\section{Analysis of network domains}
\label{analysis_of_network_domains}
In this section, we analyze the correlation profiles of the structural measures across different network domains bearing in mind the following three objectives:
\begin{enumerate}
    \item Study the structural similarities between domains 
    \item Identify domain-independent relationships between metrics \label{corr_obj_1}
    \item Choose a less redundant subset of metrics, that can still distinguish the network domains \label{corr_obj_3}
\end{enumerate}

For analyzing correlation profiles, we investigated the pairwise correlations of the metrics on correlation heatmaps and \textit{correlation networks} based on the  Spearman's rank correlation that can also measure non-linear relationships, moreover, it is less sensitive to outliers~\cite{croux2008robustness}.

\subsection{Domain similarity}
\begin{figure}
    \centering
    \includegraphics[width=0.5\linewidth]{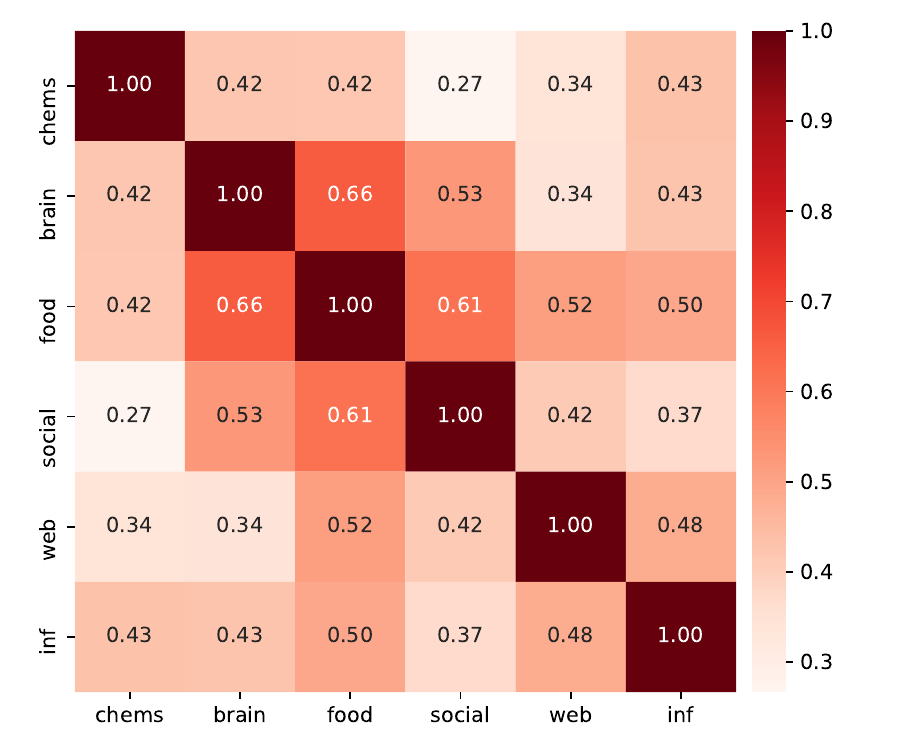}
    \caption{The similarity of the network domains based on their correlation networks. The similarities are calculated as the median adjacency similarities of correlation networks over different threshold settings.}
    \label{fig:sim}
\end{figure}
Figure \ref{fig:sim} shows the similarity of the network domains, where the similarity is captured by the pairwise Spearman's correlations of graph metrics across domains. The similarity scores are calculated through the correlation networks of the domains (using all variables detailed in Section \ref{metrics_and_notations}), namely first we created 17 correlation networks over 17 threshold settings ($0.1, 0.15, 0.2,\ldots,0.85, 0.9$) for each domain separately. Then for all thresholds, we calculated the pairwise adjacency similarities of the correlation networks of different domains. The cells of the heatmap correspond to the median of these 17 adjacency similarities. 

To calculate the adjacency similarity, we used the \texttt{similarity} function of the graph-tool package, which is the ratio of the number of edges that have the same source and target nodes in both graphs and the sum of the number of edges of the graphs. Formally let $A^{(1)}$ and $A^{(2)}$ denote the adjacency matrices of the $G_1$ and $G_2$ graphs. Then the adjacency similarity $S$ of $G_1$ and $G_2$ is defined as follows:

$$S(A^{(1)}, A^{(2)}) = \frac{\sum_{i\leq j} A_{ij}^{(1)} + A_{ij}^{(2)} - \sum_{i\leq j} \left|\,A_{ij}^{(1)} - A_{ij}^{(2)} \,\right|}{\sum_{i\leq j} A_{ij}^{(1)} + A_{ij}^{(2)}} = \frac{2\cdot \left|\, E_1 \cap E_2 \,\right|}{\left|E_1 \right| + \left|E_2 \right|},$$
where $E_1$ and $E_2$ are the edge sets of the graphs $G_1$ and $G_2$ respectively.

Figure \ref{fig:sim} suggests that the correlation profile of social, brain, and food networks are relatively similar, moreover the cheminformatics networks are rather dissimilar to the other domains. The reason behind it is that in the cheminformatics domain the correlations are relatively weak between the metrics, hence larger threshold settings shatter the correlation network into several components. Furthermore, the adjacency similarity only takes into account the mutual edges, i.e., two correlation graphs of isolated nodes are totally dissimilar. It is interesting to note that even though Ikehara \& Clauset~\shortcite{ikehara2017characterizing} applied a totally different approach, the domains that they found to be similar are consistent with our results except for the behavior of the food webs.

\subsection{Correlations of graph metrics}\label{correlation_of_metrics}
 Next, we aim to identify the strong domain-agnostic correlations among graph metrics. Here we consider the relationship of two metrics strong if their absolute Spearman's rank correlation is at least 0.65. 
 
\begin{figure}
     \centering
     \includegraphics[scale=0.6]{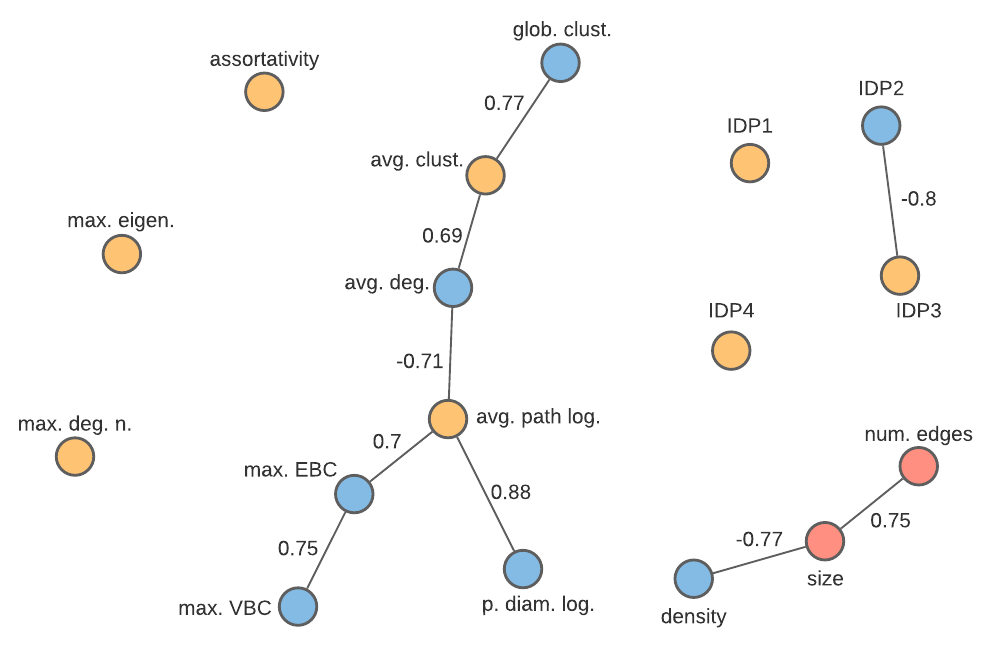}
     \caption{The correlation network of structural metrics. Two metrics are connected if the domain-averaged absolute Spearman's rank correlation is above $0.65$. The actual values of the correlations are written on the edges. The selected metrics are colored orange. Note that the max degree, avg. path length, and the pseudo diameter are normalized.}
     \label{fig:avg_abs_corr_graph}
 \end{figure}

 Figure~\ref{fig:avg_abs_corr_graph} shows the correlation network of the graph metrics, where the links are drawn if the domain-averaged absolute Spearman's rank correlation is above $0.65$. Some of the identified strong domain-agnostic relationships are not so surprising, such as the high positive correlation between the average path length and diameter, the average clustering coefficient and global clustering coefficient, and between the maximum vertex betweenness centrality and maximum edge betweenness centrality.
 
 There are also other strong, less trivial relationships among the metrics as well e.g., the strong negative correlation between density and the number of nodes, which implies that the density depends on the size of the graphs. This may be because real networks are usually sparse, i.e., the number of edges grows rather linearly than quadratically with the number of nodes, as the slope of the left scatter plot of Figure \ref{fig:size_density_avgdeg} also illustrates. On the other hand, by the definition of density, the number of edges is normalized by the square of the number of nodes. That is why considering real networks, density decays hyperbolically with the size of the graphs, as it can be seen in the middle of Figure \ref{fig:size_density_avgdeg} since the slope of the log-log plot of density against size is roughly minus one. This implies that the average degree is a less size-dependent metric than graph density (see Figure~\ref{fig:size_density_avgdeg}) which makes it a better measurement for quantifying the extent of relative connectivity of the nodes. 

The average degree and the average clustering coefficient are also positively correlated with each other regardless of domains. The reason behind this connection is that both of them measure how dense the network is, however, these are not always correlated with the graph density metric e.g., in the case of infrastructural networks. On the other hand, given the size of the network, the partial correlation of these metrics with the density is significantly larger than the Spearman's correlations. For a similar reason, there is a strong inverse connection between the average degree (and maximum degree) and the average path length, since the higher the number of edges is and the more hubs are present in the network, the shorter the average distances are. This is consistent with the finding of Bounova \& de Weck~\shortcite{bounova2012overview}, which supports that distance-related metrics (e.g., diameter and betweenness centrality) and connectivity-related metrics (e.g., average degree and clustering coefficient) are two highly connected clusters that are strongly negatively correlated, even though Bounova \& de Weck~\shortcite{bounova2012overview} investigate model-generated graphs.

Furthermore, the maximum edge betweenness centrality is always strongly positively correlated with the average path length, which can be easily seen from the definition of the two metrics. Moreover, this is again consistent with the work of Bounova \& de Weck~\shortcite{bounova2012overview} who argued that it can be analytically shown that the average betweenness centrality is linearly proportional to the normalized average path length. Similar observations are made by Jamakovic \& Uhlig~\shortcite{jamakovic2008relationships} and Garcia-Robledo \textit{et al.}~\shortcite{garcia2013correlation} despite the fact that they investigated a different set of networks.

Finally, there are universal connections between the \textit{interval degree probabilities} as well, namely strong negative correlations between IDP1 and IDP2 ($-0.61$ on average), moreover between IDP2 and IDP3 ($-0.8$ on average). This implies that IDP2 alone reasonably explains the other two variables. These relations follow from the fact that the sum of these probabilities is one, hence if one of them is relatively large, the others are necessarily low. The reason why IDP4 is usually less correlated with the other probabilities is that only 10-15\% of the degrees fall into this interval, and this percentage seems to be stable over all networks, i.e., this interval degree probability has the smallest variance. Thus, the mass of the degree distribution falls into the first three intervals, especially into the second and the third one. Due to the fact that the degree distribution of real-world networks is often heavy-tailed and asymmetric \cite{barabasi2016network}, the first interval often end up empty~\cite{aliakbary2014quantification}. The standard deviation of the degrees is so large, that there are no degrees that are smaller than the difference of the average degree and the standard deviation of the degrees, formally: \mbox{$\mu_{\text{deg}} - \sigma_{\text{deg}} < \min_v\,\deg(v)$}. Hence if IDP1 is not zero, it can only \say{steal} degrees from IDP2, that is why these two probabilities are strongly negatively correlated. On the other hand, the degrees corresponding to IDP3 are \say{closer} to the degrees of IDP2 than the degrees of IDP4, due to the heavy-tailed degree distributions again. Thus in general if IDP3 is high, IDP2 is necessarily low, that is why IDP4 is less influenced by the value of IDP3. Further study of these relationships requires a deeper understanding of the underlying domains and is beyond the scope of this paper. 

Overall, the maximum eigenvector centrality, the assortativity, and the IDP4 have the smallest correlation with the other variables. On the other hand, the average degree, average clustering coefficient, and the normalized average path length metrics have the highest correlation with the other variables.

\subsection{Metric selection}
In this section, we aim to select a smaller subset of metrics that describes the networks well but has smaller redundancy.  Using a smaller non-redundant set of metrics, the model fitting is less computationally expensive, moreover, we can also avoid collinearity and overfitting. We evaluate our selection via machine learning tools, namely by investigating the predictive power of the chosen metrics with respect to the domains of the networks. In this work, we assume that the reader is familiar with the basic concepts of machine learning, otherwise, for a detailed description of the related concepts, we refer to the book by Friedman \textit{et al.}~\shortcite{friedman2001elements}.

Based on the average absolute correlation network (Figure \ref{fig:avg_abs_corr_graph}), we select metrics from each connected component of the graph. The metrics from the component containing the size of the network are excluded from the analysis (namely the size, number of edges, and density) since the high correlation with the size implies significant trivial predictive power with respect to network domains due to the size difference of typical networks from different domains (see Figure~\ref{fig:size_density_avgdeg}). Moreover, we are more interested in finding distinctive size-independent topological properties. The selected attributes are listed in Table \ref{selected_attributes}. Our selection is slightly different from other work's~\cite{garcia2013correlation, sun2014network} since we found that skewness and kurtosis of distributions of node features are highly constrained by the size of the networks. However, assortativity, maximum degree, average path length, average clustering coefficient, and eigenvector centrality are also selected in other works~\cite{garcia2013correlation, sun2014network,harrison2014network, sala2010measurement, blasius2018towards}.

\begin{table}[h]
\centering
\begin{tabular}{ll}
\textbf{Name} & \textbf{Description} \\ \hline
Assortativity & The assortativity coefficient \\
Avg. clust. & The average local clustering coefficient \\
Avg. path log & \multirow{2}{*}{\begin{tabular}[c]{@{}l@{}}The average path length divided by the \\ logarithm of the size.\end{tabular}} \\
 &  \\
IDP's (1,3,4) & The interval degree probabilities \\
Max. deg. n. & The normalized maximum degree (max. degree centrality) \\
Max. eigen. & Maximum eigenvector centrality \\ \cline{2-2} 
Domain & \multirow{3}{*}{\begin{tabular}[c]{@{}l@{}}The domain of the real networks:\\ social, food, brain, cheminformatics (chems),\\ infrastructural and web\end{tabular}} \\
 &  \\
 &  \\
Category & \multirow{2}{*}{\begin{tabular}[c]{@{}l@{}}Indicates whether the graph is real or generated:\\ real or model\end{tabular}} \\
 &  \\
Subcategory & \multirow{5}{*}{\begin{tabular}[c]{@{}l@{}}In case of generated graphs, indicates the type\\ of the model as well: real (original), 2K, \\ CBA (clustering Barab\'asi--Albert),\\ 
FF (forest-fire),\\ 
SBM (stochastic block model).\end{tabular}} \\
 &  \\
 &  \\
 &  \\
 &  \\
 &  \\
 &  
\end{tabular}
\caption{The selected graph metrics and the other variables of the narrowed dataset.}
\label{selected_attributes}
\end{table}

In order to evaluate our metric selection, we predict the domains of the networks, first using all of the metrics, then using only the selected (numerical) attributes from Table \ref{selected_attributes}. Our selection is considered \say{optimal} if the accuracy of the predictions does not decrease significantly after we drop the redundant attributes. Table \ref{dom_pred_table} contains the average cross-validated accuracy of a few machine learning model. 
\begin{table}[]
\caption{Domain prediction performance of the machine learning models using different feature sets. The performance is calculated as the 5-fold cross validated average accuracy . The standard deviation of the accuracy is shown in brackets. We performed $1,000$ random subsampling of the attributes, with sample size 8 (since the number of selected attributes is also 8). The third and the fourth columns show the maximum and the median accuracy over all of the $1,000$ random attribute selections correspondingly. We excluded the size, number of edges and density from each attribute set.}
\label{dom_pred_table}
\centering
\begin{tabular}{l|cccc}
\multirow{2}{*}{\textbf{Classifier model}} & \multirow{2}{*}{\textbf{All attributes}} & \multirow{2}{*}{\textbf{Selected attributes}} & \multirow{2}{*}{\textbf{\begin{tabular}[c]{@{}c@{}}Maximum of\\ randomly selected\end{tabular}}} & \multirow{2}{*}{\textbf{\begin{tabular}[c]{@{}c@{}}Median of \\ randomly selected\end{tabular}}} \\
 &  &  &  &  \\ \hline
\makecell[l]{\textbf{kNN}\\} & \makecell{76.8\% \\ ($\pm$3.9\%)} & \makecell{84.4\% \\ ($\pm$1.0\%)} & \makecell{85.6\% \\($\pm$1.6\%)}  & \makecell{75.2\% \\ ($\pm$3.9\%)}\\
\makecell[l]{\textbf{Decision Tree}\\} & \makecell{82.8\% \\($\pm$1.6\%)} & \makecell{82.6\%\\ ($\pm$2.6\%)} & \makecell{84.8\% \\ ($\pm$2.3\%)} & \makecell{79.2\% \\ ($\pm$3.2\%)} \\
\textbf{Random Forest} & \makecell{87.6\% \\ ($\pm$2.2\%)} & \makecell{89.0\% \\ ($\pm$1.1\%)} & \makecell{89.4\% \\ ($\pm$2.4\%)} & \makecell{85.6\% \\ ($\pm$4.2\%)}
\end{tabular}
\end{table}
Three important observations can be drawn from Table \ref{dom_pred_table}. First, our metric selection achieves very high accuracy. Second, the kNN achieved higher accuracy when it was trained on a smaller set of features. The reason behind this is that kNN is sensitive to irrelevant features~\cite{langley1993average}, and these features bias the results. Finally, we cannot achieve significantly higher accuracy with random subsampling of the attributes, than the accuracy of our selection. Note that the best-performing random selection is usually quite similar to our selection, however, this set often contains the maximum betweenness centrality, but we would rather not include it in our selection, because the calculation of the betweenness centralities is computationally expensive.

\begin{figure}[h]
    \centering
    \includegraphics[width=0.5\linewidth]{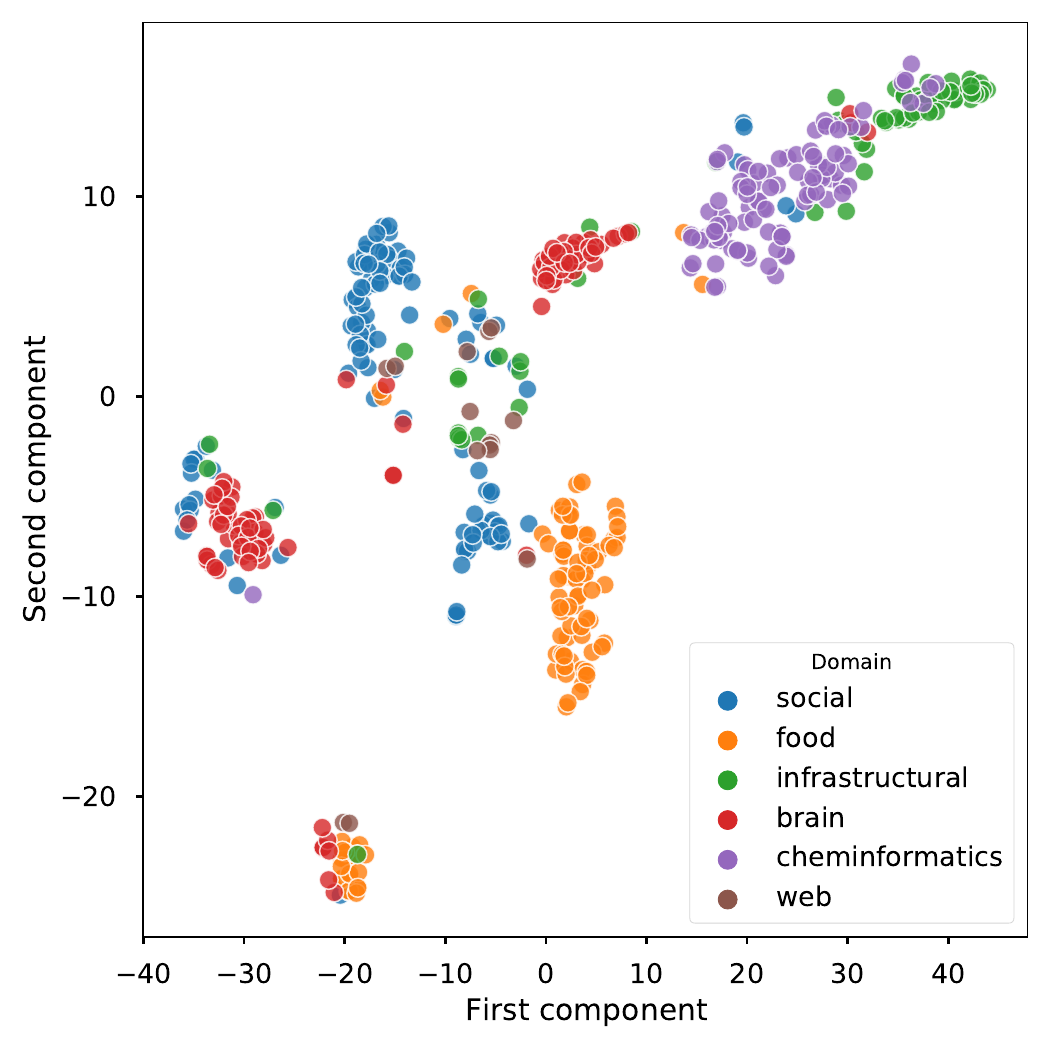}
    \caption{The t-distributed stochastic embedding (t-SNE) of the 8 selected graph metrics (detailed in Table \ref{selected_attributes}) of the real networks. Apparently, one can distinguish between the domains reasonably well using the selected metrics.}
    \label{fig:tsne}
\end{figure}

Figure \ref{fig:tsne} shows the t-distributed stochastic embedding (t-SNE)~\cite{maaten2008visualizing} of real networks represented by the selected 8 metrics. It can be seen that the network domains can be relatively clearly distinguished from each other, i.e. this figure also supports the validity of our metric selection.

\section{Analysis of network models \label{analysis_of_network_models}}
In this section, we use machine learning tools to analyze and evaluate the goodness-of-fit of the network models and the applied parameter calibration method. In particular, we solve the following two classification problems:
\begin{enumerate}
    \item \label{p1_classification} Using the selected graph metrics (Table \ref{selected_attributes}), we aim to solve a binary classification problem, where the target variable indicates whether the given network is \textit{real} or \textit{model-generated}.
    \item \label{p2_classification} Based on the selected graph metrics, we aim to identify the \textit{subcategory} of the networks. The values of subcategory attribute can be \textit{real} or the types of the models (see Table \ref{selected_attributes}). Thus, we aim to solve a multiclass classification problem that predicts whether a network is real or model-generated and if it is the latter also the model that generated it.
\end{enumerate}

Solving these classification problems allows us to answer the research objective \ref{obj_2} of this paper. Identifying the reasons why a (well-performing) machine learning model decides that a given graph is rather real than model-generated (or vica-versa) helps us find and understand the structural properties of real networks that the models are unable to capture. These decision rules are easy to retrieve for example by looking at the \textit{feature importances} of the variables. Moreover, the second classification problem helps us understand which models generate the most realistic networks and also which models generate similar graphs to each other. By carrying out the above analyses separately for each domain, we can also study which network domains are easy or difficult to synthesize. 

\begin{table}[h]
\caption{Performance of the machine learning models in the binary and multiclass classification problems. }
\label{tab:perf_ml}
\begin{tabular}{ll|rrrrrr}
\textbf{Classification problem} & \textbf{ML Model} & \multicolumn{1}{l}{\textbf{Brain}} & \multicolumn{1}{l}{\textbf{Chem.}} & \multicolumn{1}{l}{\textbf{Food}} & \multicolumn{1}{l}{\textbf{Inf.}} & \multicolumn{1}{l}{\textbf{Social}} & \multicolumn{1}{l}{\textbf{Web}} \\ \hline
\textbf{Binary classification} & kNN & 0.950 & 0.968 & 0.726 & 0.856 & 0.879 & 0.376 \\
(AUC score) & Decision Tree & 0.860 & 0.856 & 0.647 & 0.754 & 0.821 & 0.354 \\
\textbf{} & Random Forest & 0.926 & 0.970 & 0.663 & 0.885 & 0.902 & 0.164 \\ \hline
\textbf{Multiclass classification} & kNN & 89.6\% & 83.4\% & 74.8\% & 68.5\% & 77.6\% & 38.6\% \\
(Accurcacy) & Decision Tree & 89.0\% & 79.2\% & 71.2\% & 74.7\% & 83.4\% & 48.6\% \\
 & Random Forest & 91.4\% & 85.8\% & 74.0\% & 82.1\% & 83.7\% & 55.7\%
\end{tabular}
\end{table}

Table \ref{tab:perf_ml} shows the performance of the machine learning models in the binary (Problem \ref{p1_classification}) and multiclass (Problem \ref{p2_classification}) classification problems. From the table, it is apparent, that the models can relatively easily distinguish whether the graphs are real or model-generated, except for the food and especially the web domain, where both the AUC and accuracy scores are low.    

\begin{figure}[h]
    \centering
    \includegraphics[width=0.54\textwidth]{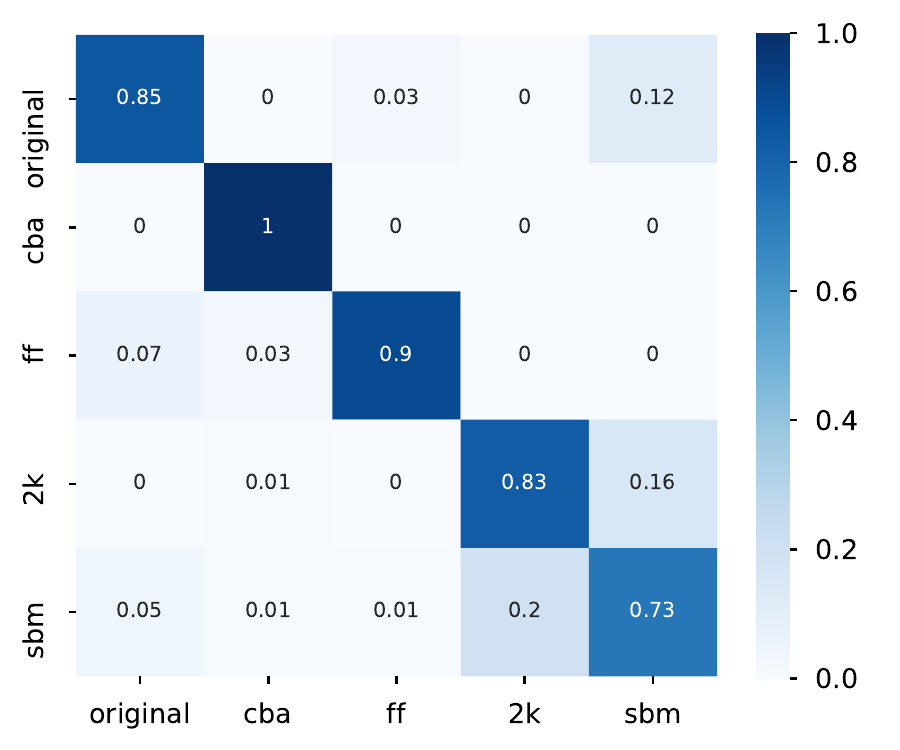}
    \includegraphics[width=0.44\textwidth]{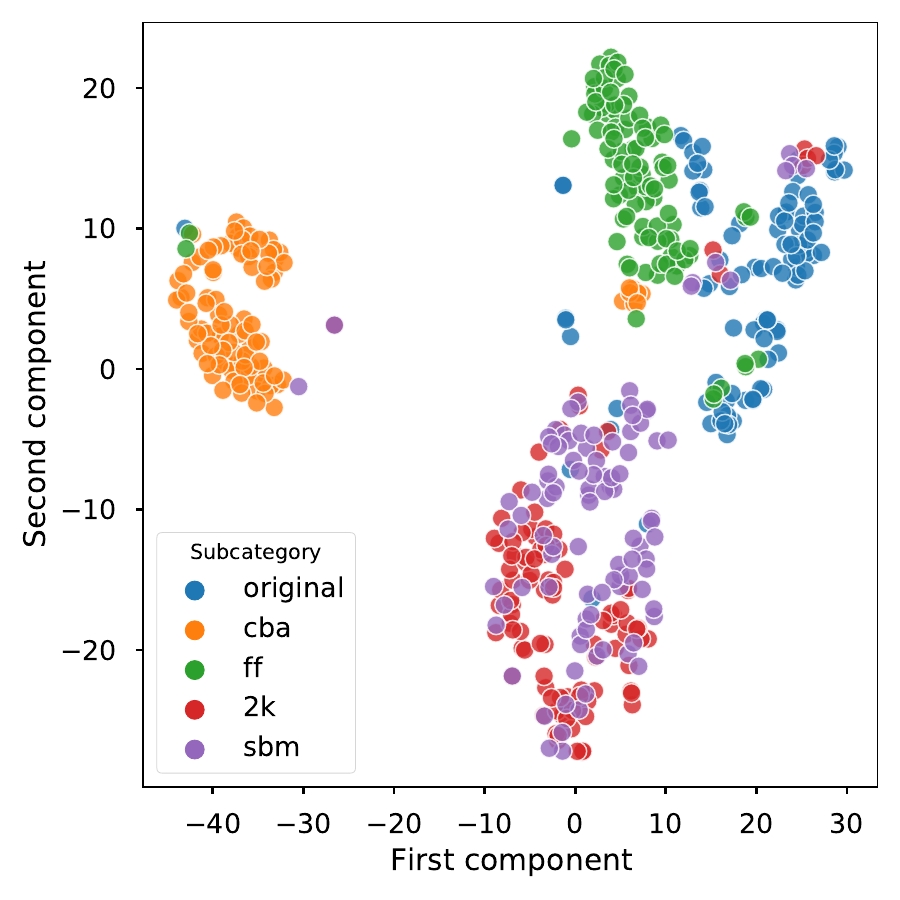}
    \caption{The normalized confusion matrix of the Random Forest on the Cheminformatics network domain (left) and the t-SNE embedding of the corresponding dataset (right). }
    \label{fig:chem_conf_tsne}
\end{figure}

\begin{figure}[h]
    \centering
    \includegraphics[width=0.54\textwidth]{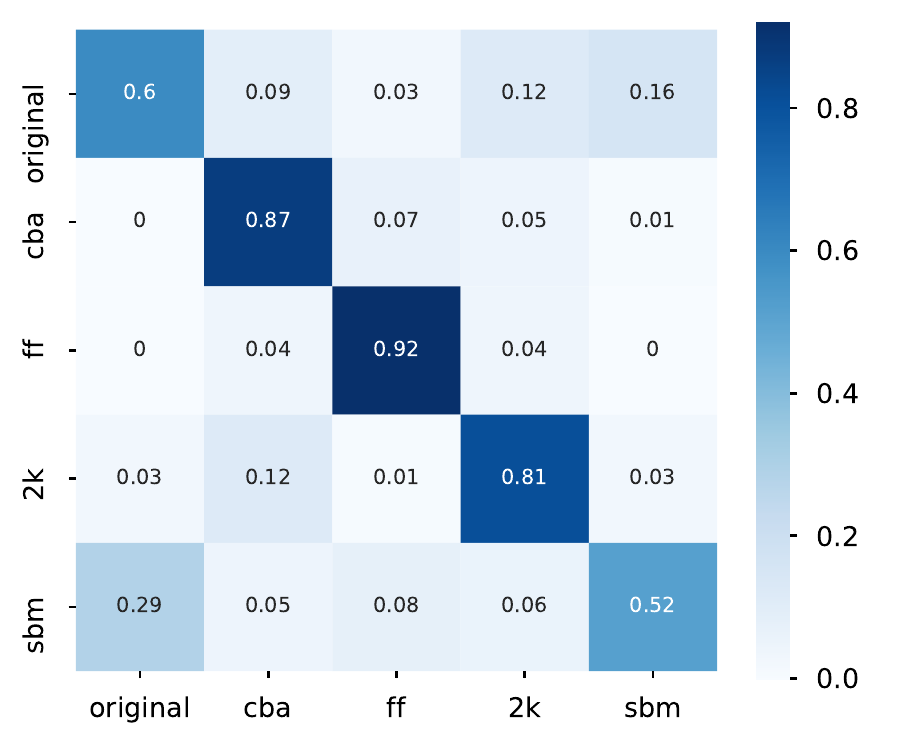}
    \includegraphics[width=0.44\textwidth]{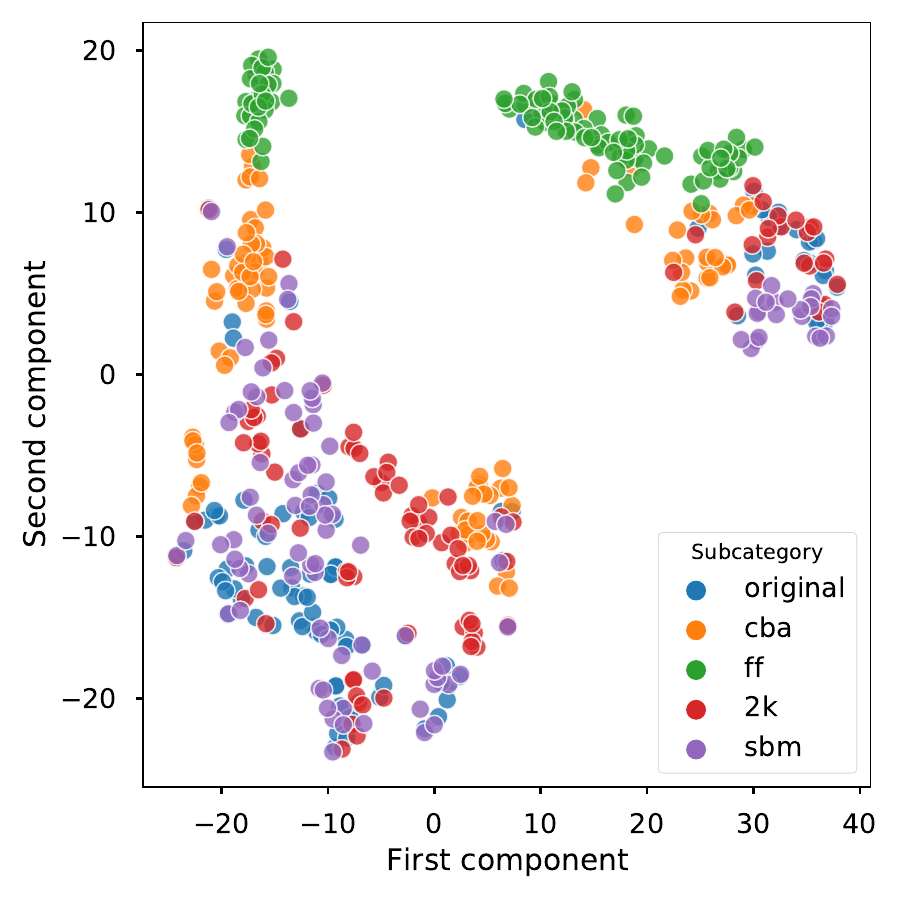}
    \caption{The normalized confusion matrix of the Random Forest on the Food network domain (left) and the t-SNE embedding of the corresponding dataset (right).}
    \label{fig:food_conf_tsne}
\end{figure}

Thus, the performance of the network models depends on the network domain. Figure~\ref{fig:food_conf_tsne} shows an example of when the network models perform relatively well, and Figure~\ref{fig:chem_conf_tsne} illustrates a counterexample, i.e. when the network models cannot really synthesize the target networks. Hence, as Figure \ref{fig:chem_conf_tsne} shows, in the cheminformatics domain the network models are not able to capture exactly the structural properties of the real networks since in the confusion matrix the accuracy for the real networks is relatively high and the real networks are rather separated in the t-SNE embedding. On the other hand, as the confusion matrix and the t-SNE embedding in Figure~\ref{fig:food_conf_tsne} illustrate, on the food domain the stochastic block model can almost perfectly mimic the networks.


According to the Random Forest classifier, when the goal is to predict whether the network is real or model-generated, the most distinguishing graph metrics are the normalized average path length, the average clustering coefficient, the maximum eigenvector centrality, and the assortativity. 

\begin{figure}
    \centering
    \includegraphics[width=0.49\textwidth]{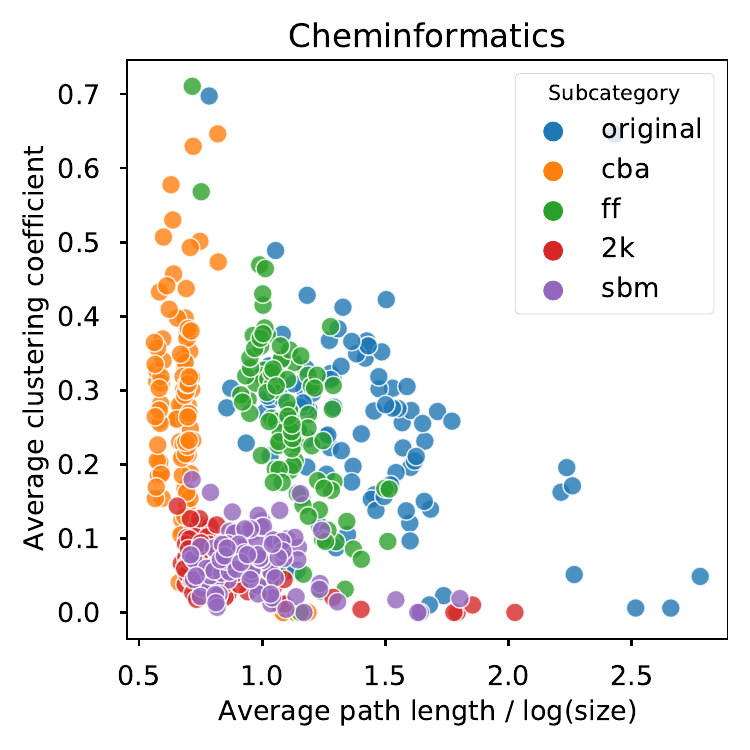}
    \includegraphics[width=0.49\textwidth]{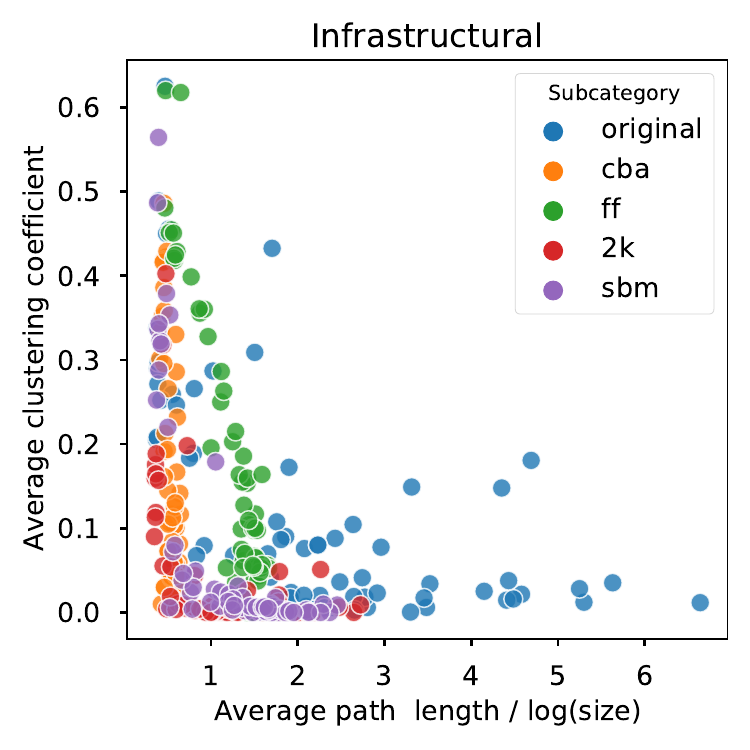}

    \includegraphics[width=0.49\textwidth]{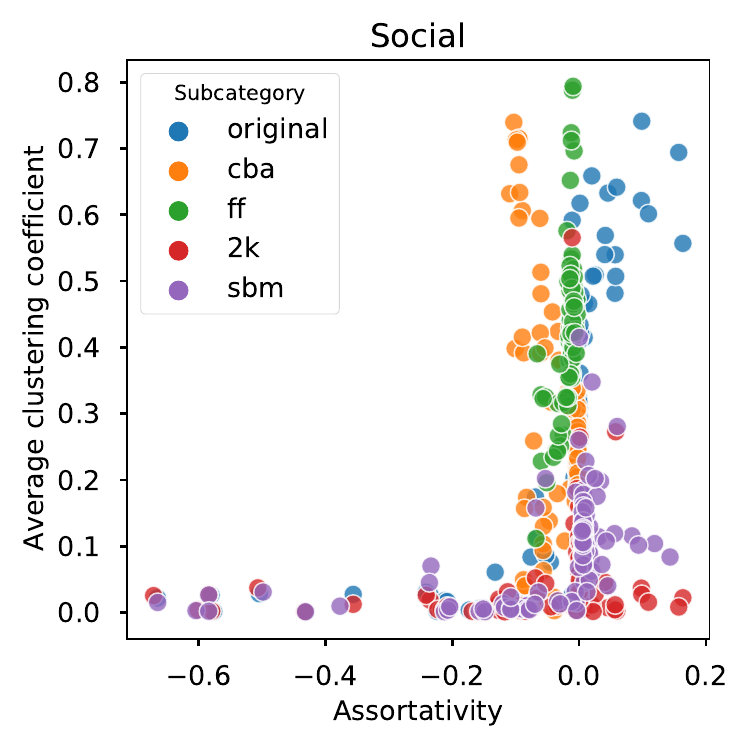}
    \includegraphics[width=0.49\textwidth]{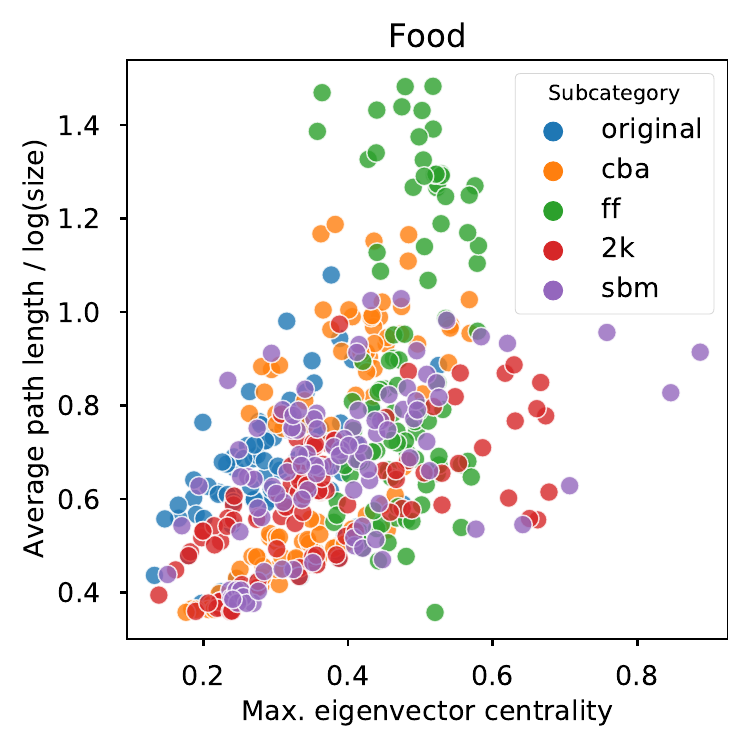}
    \caption{Distribution of the most distinguishing graph metrics of the real networks and the model-generated graphs.}
    \label{fig:difficult_prop}
\end{figure}

Figure \ref{fig:difficult_prop} illustrates the structural properties of the real networks that the network models cannot capture. For example, in the cheminformatics domain, the real graphs typically have relatively large average path length and a large clustering coefficient at the same time. However, most of the models -- except for the forest fire -- can either generate a graph with a high clustering coefficient but a small average path length or the other way around. 

Furthermore, most of the infrastructural networks have so large average path lengths that none of the models can recreate (Figure~\ref{fig:difficult_prop}). The network models usually generate small-world or ultra-small-world networks, while the diameter of a grid-like infrastructure network, for example, a road network, does not scale logarithmically.

In the case of social networks, average clustering coefficient and assortativity were the most important distinguishing metrics. As the bottom left scatterplot of Figure~\ref{fig:difficult_prop} shows, the stochastic block model and the 2K model can capture the assortativity of the real networks but cannot mimic the high average clustering coefficient. On the other hand, the clustering Barabási--Albert model and the forest-fire model can generate highly clustered networks, but the assortativity of these model-generated graphs is usually very close to zero. Hence, a subgroup of the social networks with a rather assortative and highly clustered structure cannot be synthesized with these network models. 

\begin{figure}
    \centering
    \includegraphics[width=0.49\textwidth]{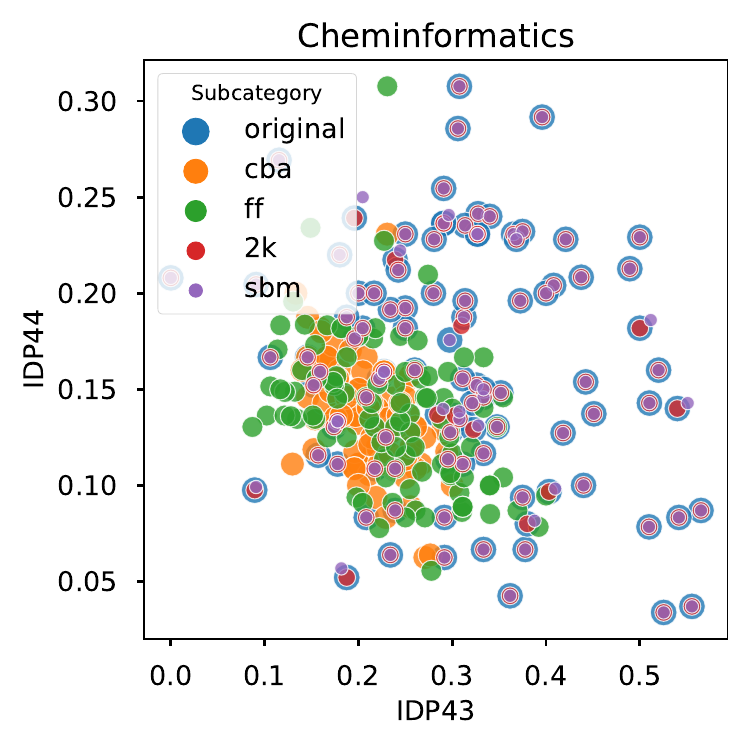}
    \includegraphics[width=0.49\textwidth]{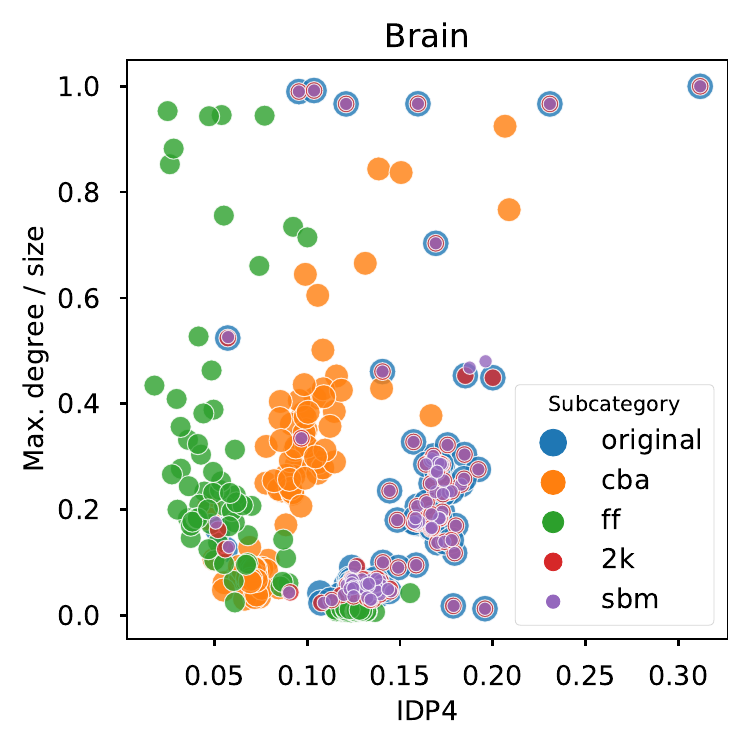}

    \includegraphics[width=0.49\textwidth]{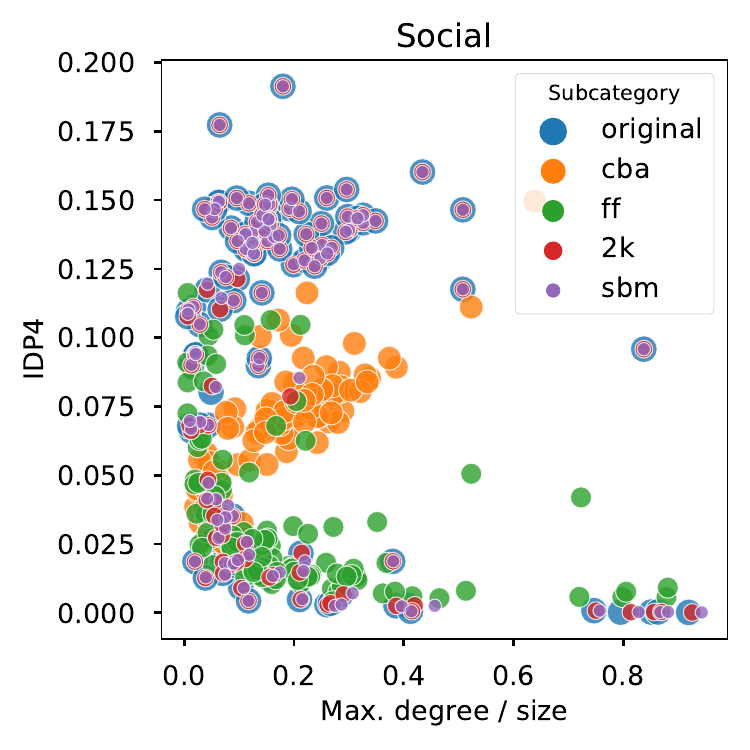}
    \includegraphics[width=0.49\textwidth]{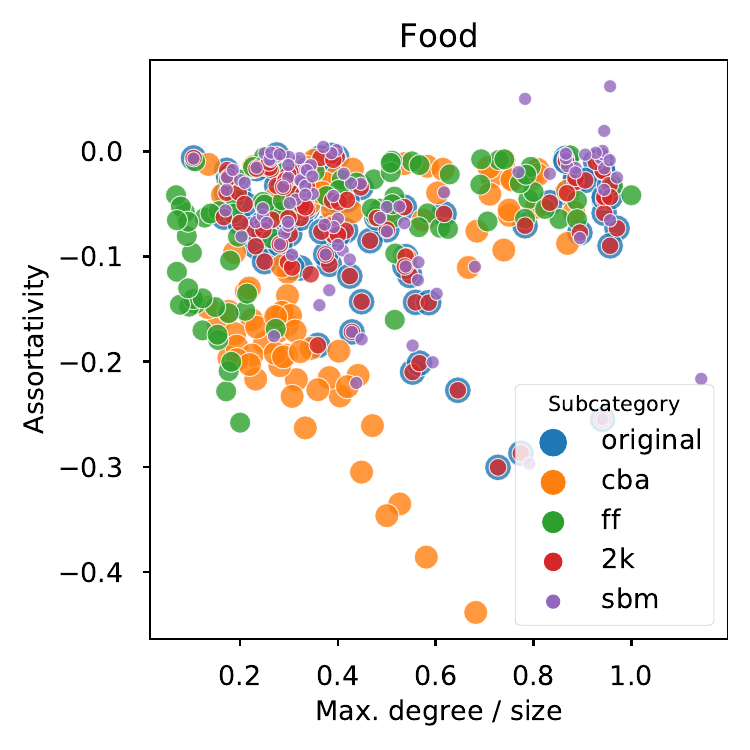}
    \caption{Distribution of the degree-distribution-related metrics of the real and the model-generated graphs. The size of the point markers varies only to be able to better see the overlapping points.}
    \label{fig:easy_prop}
\end{figure}

In the food domain, the two most distinguishing graph metrics are the maximum eigenvector centrality and the normalized average path length. The bottom-right scatter plot of Figure~\ref{fig:difficult_prop} shows that compared to the real food networks, the models either generate graphs with smaller average clustering or with larger eigenvector centrality. 

Similarly, in the case of the brain and web domains, the two most important graph metrics are average path length and the average clustering coefficient. The scatter plots for these domains can be seen in the supplementary material \cite{supplementary}.

The models, especially the 2K and the stochastic block model, can rather efficiently capture the other metrics, namely the normalized maximum degree and the interval degree probabilities. Figure \ref{fig:easy_prop} shows that in the majority of the cases, not so surprisingly the 2K and stochastic block model can exactly capture the aforementioned degree-distribution-related metrics, that is why the corresponding points on the scatter plots completely overlap each other. However, a less sophisticated model, the forest-fire model can also achieve relatively high accuracy regarding these graph metrics, as the bottom-left scatter plot of Figure~\ref{fig:easy_prop} depicts.

\begin{figure}
    \centering
    \includegraphics[width=0.325\textwidth]{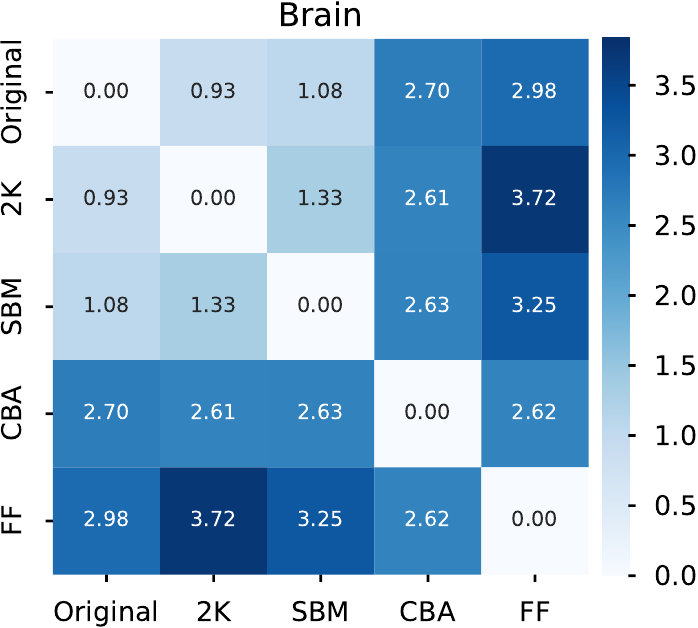}
    \includegraphics[width=0.325\textwidth]{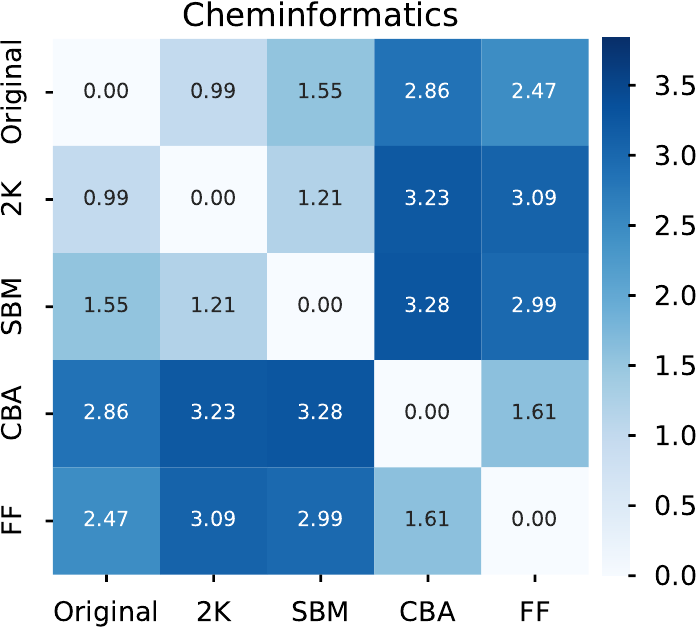}
    \includegraphics[width=0.325\textwidth]{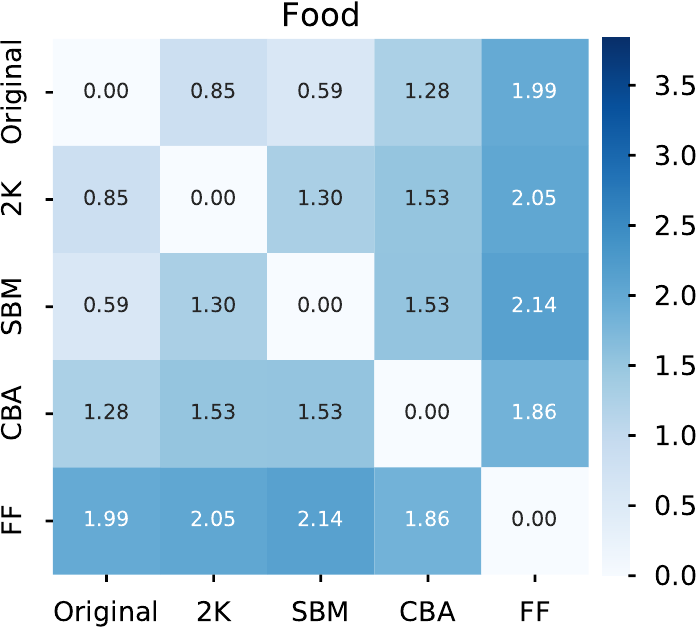}
    \includegraphics[width=0.325\textwidth]{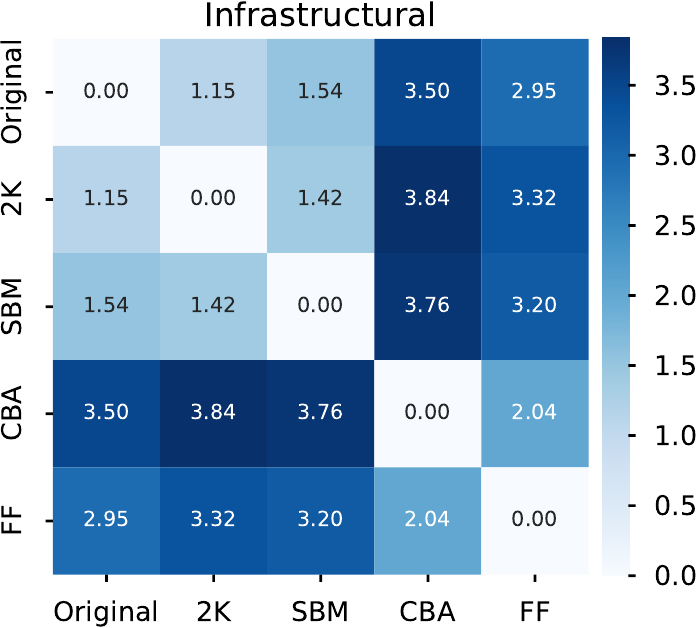}
    \includegraphics[width=0.325\textwidth]{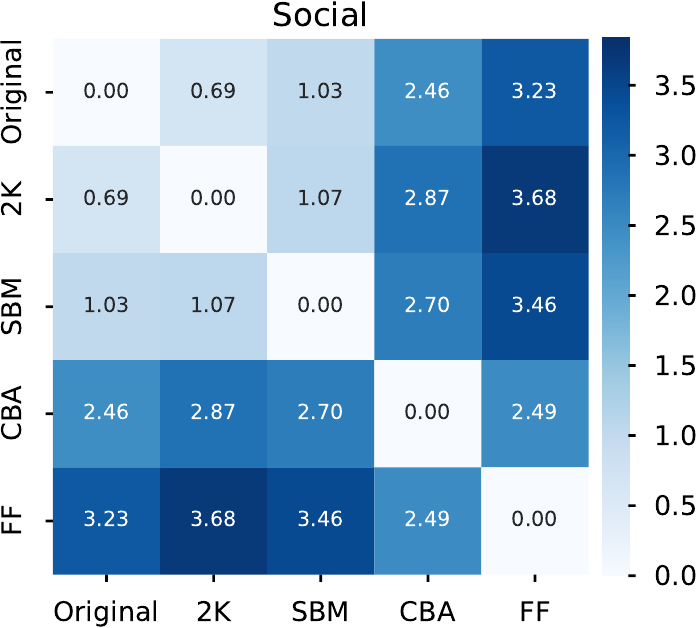}
    \includegraphics[width=0.325\textwidth]{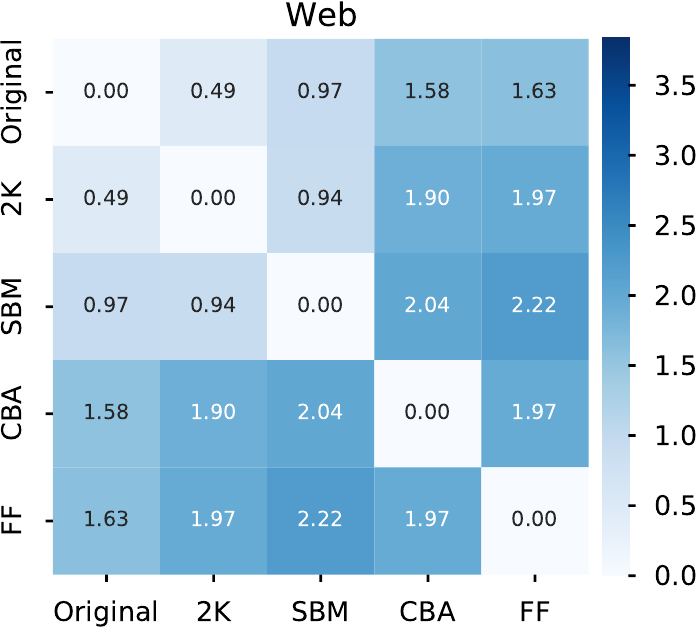}
    \caption{The mean Canberra distance between the real and the corresponding model-generated graphs.}
    \label{fig:mean_can_dist}
\end{figure}

Figure \ref{fig:mean_can_dist} shows the mean Canberra distance between the (selected) metrics of the real target graphs and the fitted model-generated graphs. From the figure, it is apparent that the 2K and the stochastic block models are the most accurate network models in all domains. Moreover, the graphs generated by the forest-fire are usually closer to the graphs of the clustering Barabási--Albert model than to the real networks. This may be due to the fact that these two models have a similar growing mechanism. Finally, Figure~\ref{fig:easy_prop} also indicates that the food and the web domains are the easiest to mimic with these network models, which is also in alignment with Table \ref{tab:perf_ml} since the distinguishing ability of the machine learning models was the lowest on these domains. 

\section{Conclusion \label{conclusion}}
In this paper, we applied and unified related but so far rather separated branches of data-driven analysis of real and model-generated networks, such as the study of graph metrics, network similarity, network model calibration, and graph classification. First, after a thorough and critical survey of the related state-of-the-art approaches of the literature, we constructed a large dataset to investigate the characteristics and structural properties of real-world networks from six domains and the model-generated counterparts.

By studying the correlation profile of the graph metrics of the real networks, we identified a small, uncorrelated subset of metrics that efficiently describe the real networks. Using the selected nonredundant graph metrics, we fitted each network model to each real network, i.e., we calibrated the models' parameters such that they are as close to the real networks as possible. Note that, we also showed that the distribution of the graph metrics of the random network models is concentrated enough to be able to use them as a base to perform parameter calibration. To be able to measure the similarity of the graphs we used the Canberra distance of the vectors of the selected metrics. 

With the help of machine learning techniques, we found that the network models are unable to capture the highly clustered structure and relatively large average path length of the real-world networks. On the other hand, the models are able to capture the degree-distribution-related metrics such as the degree centrality, and interval degree probabilities.  We also found that the networks from different domains are not equally modelable, for example, the brain and cheminformatics networks are the most difficult and the food and web networks are the easiest to synthesize. 

\section*{Funding Statement}
The research reported in this paper is part of project no.~BME-NVA-02, implemented with the support
provided by the Ministry of Innovation and Technology of Hungary from the National Research,
Development and Innovation Fund, financed under the TKP2021 funding scheme.
The work of R. Molontay is supported by the NKFIH K123782 research grant.

\section*{Competing Interest Statement}
Competing Interests: None

\section*{Data Availability Statement}
The datasets analyzed in this study are publicly available at the following GitHub repository: \url{https://github.com/marcessz/Complex-Networks}.

\bibliographystyle{nws}
\bibliography{references}
\end{document}